\documentclass[%
 reprint,
 amsmath,amssymb,aps,
]{revtex4-2}

\usepackage{graphicx}
\usepackage{dcolumn} 
\usepackage{bm} 
\usepackage{xcolor}

\begin{document}


\title{Doped two-dimensional diamond: properties and potential applications}

\author{Bruno Ipaves}
\email{ipaves@if.usp.br}
 \affiliation{Instituto de Física, Universidade de São Paulo, CEP 05508-090 S\~ao Paulo, SP, Brazil} 
 \author{João F. Justo}
 \email{joao.justo@usp.br}
\affiliation{Escola Politécnica, Universidade de São Paulo, CEP 05508-010, São Paulo, SP, Brazil}
\author{Biplab Sanyal}
 \email{biplab.sanyal@physics.uu.se}
\affiliation{Department of Physics and Astronomy, Uppsala University, 75120 Uppsala, Sweden}
\author{Lucy V. C. Assali}
 \email{lassali@if.usp.br}
\affiliation{Instituto de Física, Universidade de São Paulo, CEP 05508-090 São Paulo, SP, Brazil}

\date{\today}
 
\begin{abstract}
This paper examines the structural, thermodynamic, dynamic, elastic, and electronic properties of doped 2D diamond C$_4$X$_2$ (X = B or N) nanosheets in both AA$'$A$''$ and ABC stacking configurations, by first-principles calculations. Those systems consist of three diamond-like graphene sheets,  with an undoped graphene layer between two 50\% doped ones. Our results,  based on the analysis of {\it ab-initio} molecular dynamics simulations, phonon dispersion spectra, and Born's criteria for mechanical stability,  revealed that all four structures are stable. Additionally, their standard enthalpy of formation values are similar to the one of pristine 2D diamonds, recently synthesized by compressing three graphene layers together. The C$_4$X$_2$ (X = B or N) systems exhibit high elastic constant values and stiffness comparable to the diamond. The C$_4$N$_2$ nanosheets present wide indirect band gaps that could be advantageous for applications similar to the ones of the hexagonal boron nitride (h-BN), such as a substrate for high-mobility 2D devices. On the other hand,  the C$_4$B$_2$ systems are semiconductors with direct band gaps, in the 1.6 - 2.0 eV range, and small effective masses, which are favorable characteristics to high carrier mobility and optoelectronics applications.
\end{abstract}


\maketitle


\section{INTRODUCTION}
\label{sec:INTRODUCTION}

Graphene is the most popular two-dimensional (2D) material, being a zero-gap semimetal with a honeycomb carbon structure and $sp^2$ hybridization. It carries a unique combination of physical properties in nature, such as high electrical conductivity, tensile strength, and optical transparency. Additionally, it is the elementary structure for several other nanomaterials, such as fullerenes, nanotubes, graphite, and the single-layer diamond (2D diamond) \cite{novoselov2004, geim2007rise, gao2018ultrahard}. As a result of recent developments in the synthesis and characterization of 2D materials, the 2D diamond has received great attention, with promising applications in several fields, such as batteries, quantum computing, nano-optics, and nanoelectronics \cite{sorokin2021two}.

The stabilization of 2D diamond often requires surface functionalization, leading to a variety of structures, which have received different labels, such as diamane, diamene, diamondol, and diamondene \cite{sorokin2021two, qin2021diamane}. 2D diamonds can also be built out of bilayer graphene (BLG) or few-layer graphene (FLG) through different techniques. For example, the hydrogenated (HD) and fluorinated (FD) 2D diamonds can be synthesized at ambient pressure without a substrate, in which the HD can be produced using hot filament chemical vapor deposition (CVD) \cite{piazza2019low}, while FD by combining FLG and gaseous ClF$_3$ \cite{grayfer2013synthesis}.

The pristine 2D diamond (PD) is hard to synthesize as high pressures are required to transform $sp^2$ bonds from graphene layers into interlayer $sp^3$ ones \cite{qin2021diamane}. Nevertheless, the PD has recently been synthesized without a substrate, by compressing three graphene layers \cite{ke2020synthesis}. Additionally, a theoretical investigation has shown that it is possible to stabilize the 2D diamond made of two graphene layers with nitrogen substitution \cite{pakornchote2020roles}. For example, the NCCN 2D structure, composed of two carbon layers functionalized with nitrogen ones on both sides, has also been investigated, suggesting it could be used as a selective ammonia sensor \cite{ipaves2019carbon, he2020electric, tan2020novel}. 

The physical properties of 2D diamonds may vary considerably, depending on the synthesis methods, leading to structures with different configurations, functional groups, and heteroatoms \cite{qin2021diamane, sorokin2021two}. At room temperature, the thermal conductivity of HD is high and the heat transport arises from the acoustic phonon modes. On the other hand, under the same conditions, the thermal conductivity of FD is lower than that of HD and the heat transport is controlled by the optical phonon modes \cite{zhu2019suppressed}. 2D diamonds also present remarkable mechanical properties, with stiffness and Young’s modulus similar to the ones of graphene and bulk diamond \cite{cellini2018epitaxial}. Furthermore, unlike graphene, 2D diamonds have band gap features that depend on the stacking arrangement, the number of layers, and the functional groups present in the structures \cite{cheng2020high, sorokin2021two}. Despite several recent experimental and theoretical investigations on 2D diamonds, the origin of all these peculiar properties has been the subject of debate \cite{qin2021diamane, sorokin2021two}. 

In this paper, we present a study of the physical properties of 2D diamonds doped with substitutional N or B atoms. The reference systems consist of three graphene sheets: an undoped graphene layer between two 50\% doped ones, where the C-C bonds between neighboring layers are strong covalent bonds. Here, we considered four structure configurations labeled AA$'$A$''$-C$_4$N$_2$, ABC-C$_4$N$_2$, AA$'$A$''$-C$_4$B$_2$, and ABC-C$_4$B$_2$. Their structural, thermodynamic, dynamic, elastic, and electronic properties and potential applications are explored and discussed in depth. 

\section{COMPUTATIONAL METHODS}
\label{sec: COMPUTATIONAL METHODS}

This investigation was performed using first-principles calculations based on the Density Functional Theory (DFT) \cite{kohn1965self}, using the plane-wave basis set and projector augmented-wave (PAW) method \cite{kresse1999ultrasoft}, as implemented in the Quantum ESPRESSO computational package \cite{giannozzi2009quantum,giannozzi2017advanced}. We utilized the generalized gradient approximation of Perdew–Burke–Ernzerhof (GGA-PBE) exchange–correlation functional \cite{perdew1996generalized} and the Dion {\it et al.} scheme \cite{dion2004} optimized by Klime{\v{s}} {\it et al.} (optB88-vdW \cite{klimevs2009}) to properly describe the effects of the dispersive van der Waals (vdW) interactions. For an accurate description of the energy band gap values, we employed the hybrid Heyd-Scuseria-Ernzerhorf (HSE) functional \cite{heyd2003hybrid} at the relaxed structures obtained from the optB88-vdW approximation. The plane-wave energy cutoff was set to 1100 eV with a convergence threshold of 0.1 meV/atom for the total energy. We used a $16 \times 16 \times 1$ $k$-point mesh to describe the irreducible Brillouin zone \cite{monkhorst1976special}, and the forces on atoms were converged down to 1 meV/{\AA}. To obtain the phonon dispersion curves, we used the Density Functional Perturbation Theory (DFPT) \cite{baroni2001phonons} with an $8 \times 8 \times 1$ $q$-point mesh. 

The primitive hexagonal cells of the 2D structures were constructed using 6 atoms. To determine the cell parameters in the $xy$-plane, a variable-cell optimization was carried out with the BFGS quasi-newton algorithm. In order to avoid interactions among cell images, the lattice parameter perpendicular to the sheets ($z$-axis) was fixed at 25 Å. This approach has been successfully applied to similar 2D systems in previous studies \cite{garcia2011group, ipaves2019carbon, ipaves2023aluminum}.

In order to determine the elastic properties of the systems, we built a rectangular cell with 12 atoms and used the strain–energy method \cite{cadelano2010elastic, ipaves2022functionalized}. Accordingly, for isotropic structures and small deformations $(\epsilon)$ near their equilibrium configurations, the elastic energy, per unit area, was approximated as
\begin{equation}
 E(\epsilon) - E(0) \approx \frac{1}{2}E^{(2)} \epsilon^{2},
\label{eq_elastic_energy}
\end{equation}
where $E(\epsilon)$ is the total energy of strained con\-fig\-u\-ra\-tions, while $E(0)$ is the total energy of the respective unstrained ones. We applied two in-plane deformations, ranging from -1.2\% to 1.2\%, in order to obtain the $E^{(2)}$, which allowed to obtain the elastic constants after fitting a second-order polynomial to the data. Herein, $E^{(2)} = C_{11}$ elastic constant for the zigzag axial deformation, while $E^{(2)} = 2(C_{11} + C_{12})$ for the biaxial planar deformation \cite{cadelano2010elastic,ipaves2022functionalized}.

The thermal stability was studied by computing the standard enthalpy of formation, per atom, of the structures at 0 GPa ($\Delta H_{\!f}^0$), by using 
\begin{equation}
 \Delta H_{\!f}^0 = \frac{E_t({\rm C}_4{\rm X}_2) - 4E_t({\rm C}) - 2E_t({\rm X})}{6},
 \label{eq_formation_energy}
 \end{equation}
where $E_t({\rm C}_4{\rm X}_2)$ is the total energy of the 2D nanosheet, with 4 C atoms and 2 X atoms (X = B or N) in the primitive cell. $E_t$(C) and $E_t$(X) are the total energies, per atom, of the respective C and X standard ground states, i.e.,  of graphite and the crystalline boron in the trigonal structure ($\beta$-boron) or the isolated N$_2$ molecule. This procedure to determine enthalpies and/or energies of formation has been successfully used to investigate several other systems \cite{larico2004,assali2006manganese,assali20113,haastrup2018computational,marcondes2021}. 

Additionally, {\it ab-initio} molecular dynamics simulations (AIMD) were carried out using the Vienna {\it ab initio} simulation package (VASP) \cite{kresse1996efficient}, where a 6 $\times$ 6 $\times$ 1 hexagonal 216-atom supercell was adopted to allow possible structural reconstructions. A Nose–Hoover thermostat (NVT) ensemble was employed, from 300 to 1000 K for 5 ps, with a simulation time step of 1 fs.

\section{RESULTS AND DISCUSSION}
\label{RESULTS}

Initially, we explored the physical properties of pristine 2D diamond in AA$'$A$''$ and ABC stacking structural configurations, composed of three graphene layers in which the C atoms between layers are covalently bonded with a near $sp^3$ hybridization. Starting the simulations with the diamond-like configuration, they converged, after the relaxation of atomic positions, to trilayer graphene systems with vdW interactions between layers (graphite-like). This behavior has also been found in a previous theoretical investigation of 2D diamond, starting the simulations with two graphene layers \cite{pakornchote2019phase}. These results can be understood as a consequence of the absence of an external pressure-induced and/or surface passivation to promote the $sp^2$ to $sp^3$ hybridization transformation \cite{pakornchote2019phase}. Those pristine structures represent the reference systems used here to study and understand the effects of their functionalization. 
 
\begin{figure}[hbt]
\centering
\includegraphics[scale = 0.14, trim={0cm 0cm 0cm 0cm}, clip]{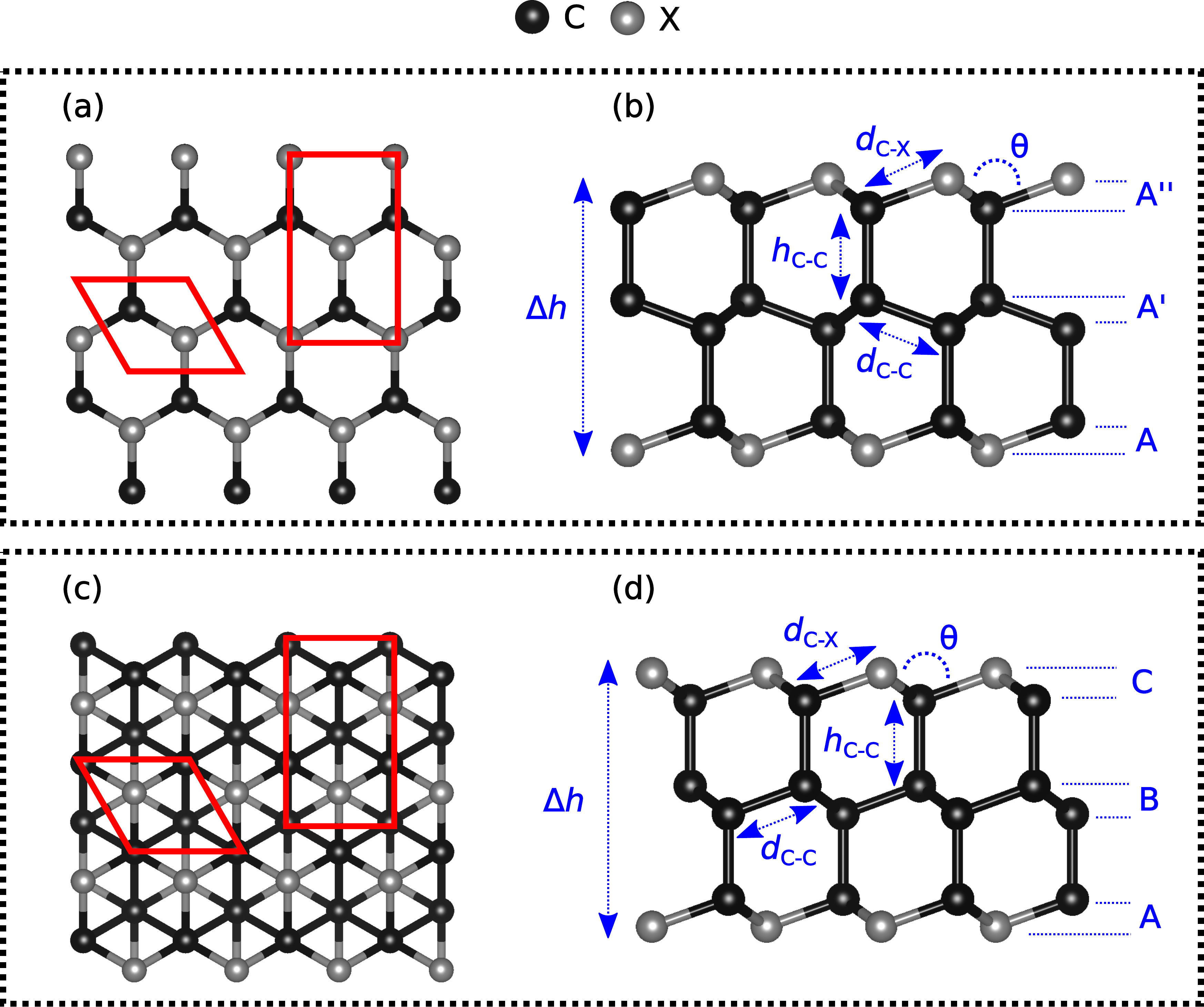} \quad
\caption{Schematic illustration of the C$_4$X$_2$ (X = B or N) systems. (a) Top and (b) side views of AA$'$A$''$-C$_4$X$_2$, (c) top and (d) side views of ABC-C$_4$X$_2$. The black and gray spheres represent the C and X atoms, respectively. The red lines denote the simulation unit cell limits, with the rectangle cells used to determine the elastic properties. The graphs also indicate the labels given to the intralayer ($d_{\rm C-X}$) and interlayer ($h_{\rm C-X}$) distances, structure thickness ($\Delta h $), and the intraplanar bond angle ($\theta$).}
\label{schematic_representation}
\end{figure}

 \begin{table}[hbt]
\caption{Structural properties of C$_4$X$_2$ (X = B or N): lattice parameter ($a$), intralayer ($d$) and interlayer ($h$) distances, thickness ($\Delta h$), and the intraplanar bond angle ($\theta$), labeled according to figure \ref{schematic_representation}. The distances are given in {\AA} and angles in degrees. The standard enthalpies of formation ($\Delta H_{\!f}^{0}$) at 0 GPa are given in meV/atom. For PD, HD, and FD, X = C.}
\begin{ruledtabular}
\begin{tabular}{lccccccc}
\textrm{System} & 
\textrm{$\!\!a$} & \textrm{$\!\!d_{\mathrm{C-X}}$}  &  \textrm{$d_{\mathrm{C-C}}$}  & \textrm{$\!\!h_{\mathrm{C-C}}$} & \textrm{$\!\!\Delta h$} & \textrm{$\!\!\theta $} & $\Delta H_{\!f}^{0}$   \\
& & & & & & & \\[-3mm]
\hline
& & & & & & & \\[-2.7mm]
AA$'$A$''$-C$_4$N$_2$ & $\!\!$2.42 & $\!\!$1.49 &  1.49 &  $\!\!$1.60  &  $\!\!$4.74  & $\!\!$108.9 & $\!\!$348\\
ABC-C$_4$N$_2$ &  $\!\!$2.44 & $\!\!$1.50 & 1.50 & $\!\!$1.57  &  $\!\!$4.66 & $\!\!$109.0 & $\!\!$333 \\
& & & & & & & \\[-2.5mm]
AA$'$A$''$-C$_4$B$_2$ & $\!\!$2.66 & $\!\!$1.55 & 1.62  &  $\!\!$1.66  & $\!\!$4.25   & $\!\!$118.4 & $\!\!$424 \\
ABC-C$_4$B$_2$ &  $\!\!$2.67 & $\!\!$1.55   & 1.63 & $\!\!$1.65 & $\!\!$4.18  & $\!\!$118.5  & $\!\!$365 \\
& & & & & & & \\[-2.5mm]
PD\footnote{Reference \cite{ke2020synthesis}. The $(\overline{2}110)$-oriented h-diamane exhibits two \textrm{$d_{\mathrm{C-X}}$} and \textrm{$\theta$ values}. Y =  bond lengths are 1.35 and 1.54 {\AA} with the angles presenting $sp^3$ and $sp^2$ hybridizations.}
& $\!\!$2.43 & Y$^{\text{a}}$ & 1.54 &  $\!\!$1.65 &  $\!\!$--- & Y$^{\text{a}}$ & $\!\!$300\\  
HD\footnote{Reference \cite{cheng2020high}} & $\!\!$2.53 & $\!\!$1.56 &  --- & $\!\!$1.56 & $\!\!$--- & $\!\!$--- & $\!\!$--- \\
FD$^{\text{b}}$ & $\!\!$2.56 & $\!\!$1.55 &  --- & $\!\!$1.55 &  $\!\!$--- & $\!\!$--- & $\!\!$--- \\
NCCN & 2.39\footnote{Reference \cite{ipaves2019carbon} }  &   1.47$^{\text{c}}$  & ---  & 1.58$^{\text{c}}$  & 2.59\footnote{Reference \cite{pakornchote2020roles}} & 108.8$^{\text{c}}$ & 211$^{\text{d}}$ \\ 
\end{tabular}
\label{tab:table_structural} 
\end{ruledtabular}

\end{table}

\begin{figure*}[hbt]
\centering
\includegraphics[scale = 0.23, trim={0cm 0cm 0cm 0cm}, clip]{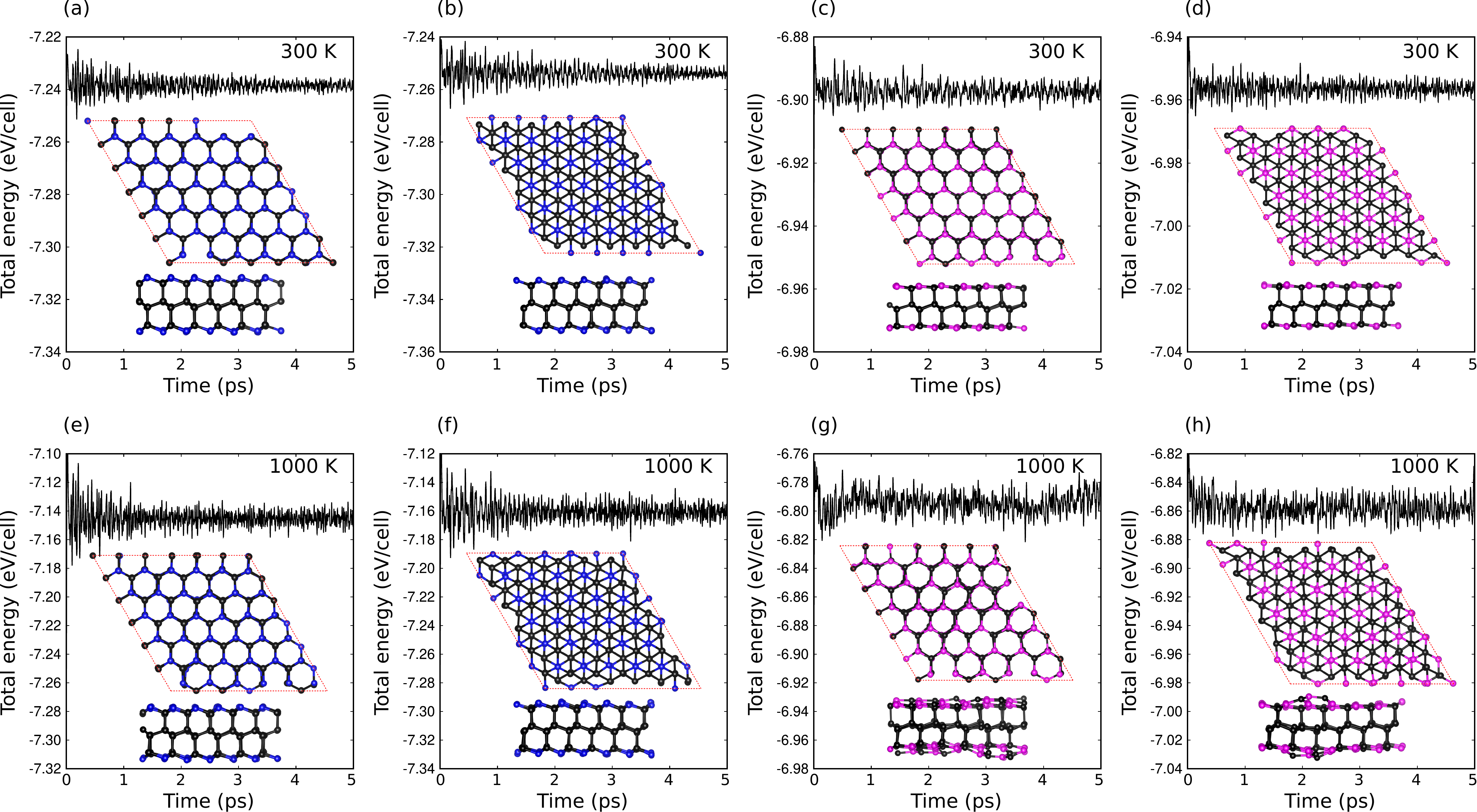} \quad
\caption{Systems' total energy variation during 5 ps, computed with AIMD simulations. At 300 K, results of (a) AA$'$A$''$-C$_4$N$_2$, (b) ABC-C$_4$N$_2$, (c) AA$'$A$''$-C$_4$B$_2$, and (d) ABC-C$_4$B$_2$. At 1000 K, results of (e) AA$'$A$''$-C$_4$N$_2$, (f) ABC-C$_4$N$_2$, (g) AA$'$A$''$-C$_4$B$_2$, and (h) ABC-C$_4$B$_2$.}
\label{figure_aimd}
\end{figure*}

Then, we explored the properties of C$_4$X$_2$ (X = B or N) systems, which can be described as three graphene sheets in which four C atoms are bonded covalently (2D diamond-like) in each unit cell. The two external layers are 50\% doped with substitutional X atoms, hence, each X atom is bonded to three C atoms. Figure \ref{schematic_representation} presents a schematic representation of the optimized and relaxed C$_4$X$_2$ (X = B or N) systems, in both AA$'$A$''$ and ABC stacking configurations, as well as the respective labels given to the intraplanar bond angle ($\theta$), intralayer ($d_{\rm C-X}$) and interlayer ($h_{\rm C-X}$) distances, and systems' thickness ($\Delta h$).
 
The optimized structural parameters of C$_4$X$_2$ (X = B or N) nanosheets are shown in Table \ref{tab:table_structural}, where the distance labels are consistent with the ones defined in figure \ref{schematic_representation}. It can be observed that all the nanosystems functionalized with N atoms keep the lattice constants almost unchanged when compared to the PD ones. Additionally, for both stacking configurations of the C$_4$N$_2$, the intraplanar bond angle ($\theta$) values are close to the $sp^3$ hybridization ones (109.47$^{\circ}$), leading to a thickness of $\approx$ 4.7 {\AA}. Nevertheless, the C$_4$B$_2$ nanosheet lattice parameters are slightly greater than those of HD and FD systems, with $\theta$ close to the value of 120$^{\circ}$, i.e., the B atoms bonded to three adjacent C atoms present a $sp^2$-type hybridization and hence we observed a smaller thickness of $\approx$ 4.2 {\AA} as compared to the N-functionalized structures.

We now discuss the stability of the C$_4$X$_2$ nanosheets. To study the thermal stability of those structures, we computed the standard enthalpy of formation $\Delta H_{f}^{0}$ using equation (\ref{eq_formation_energy}). Herein, we found positive $\Delta H_{f}^{0}$ of 424, 365, 348, and 333 meV for AA$'$A$''$-C$_4$B$_2$, ABC-C$_4$B$_2$, AA$'$A$''$-C$_4$N$_2$, and ABC-C$_4$N$_2$, respectively, displayed in Table \ref{tab:table_structural}, indicating possible thermodynamic instability. However, the literature has reported 2D materials synthesized by endothermic processes ($\Delta H_{f}^{0} > 0$), such as graphene, germanene, and silicene \cite{haastrup2018computational, ipaves2022functionalized}. Also, the $\Delta H_{f}^{0}$ values of C$_4$X$_2$ nanosheets are similar to the 300 meV/atom of PD with three graphene layers at 0 GPa (Table \ref{tab:table_structural}), which was recently synthesized \cite{ke2020synthesis}, and slightly higher than NCCN that was theoretically studied to stabilize the 2D diamond without any passivation \cite{pakornchote2020roles}.

The thermodynamic stability of the systems was also investigated by AIMD simulations. The results exhibited a total energy small variation during 5 ps at 300 K, as shown in figure \ref{figure_aimd}, indicating that the structural integrity of those systems is maintained at those conditions. At 1000 K, the same behavior is observed for the C$_4$N$_2$ systems, while the C$_4$B$_2$ nanosheets presented some broken bonds, suggesting some structural degradation.

\begin{figure}[hbt]
\centering
\includegraphics[scale = 0.31, trim={0cm 0.0cm 0cm 0.0cm}, clip]{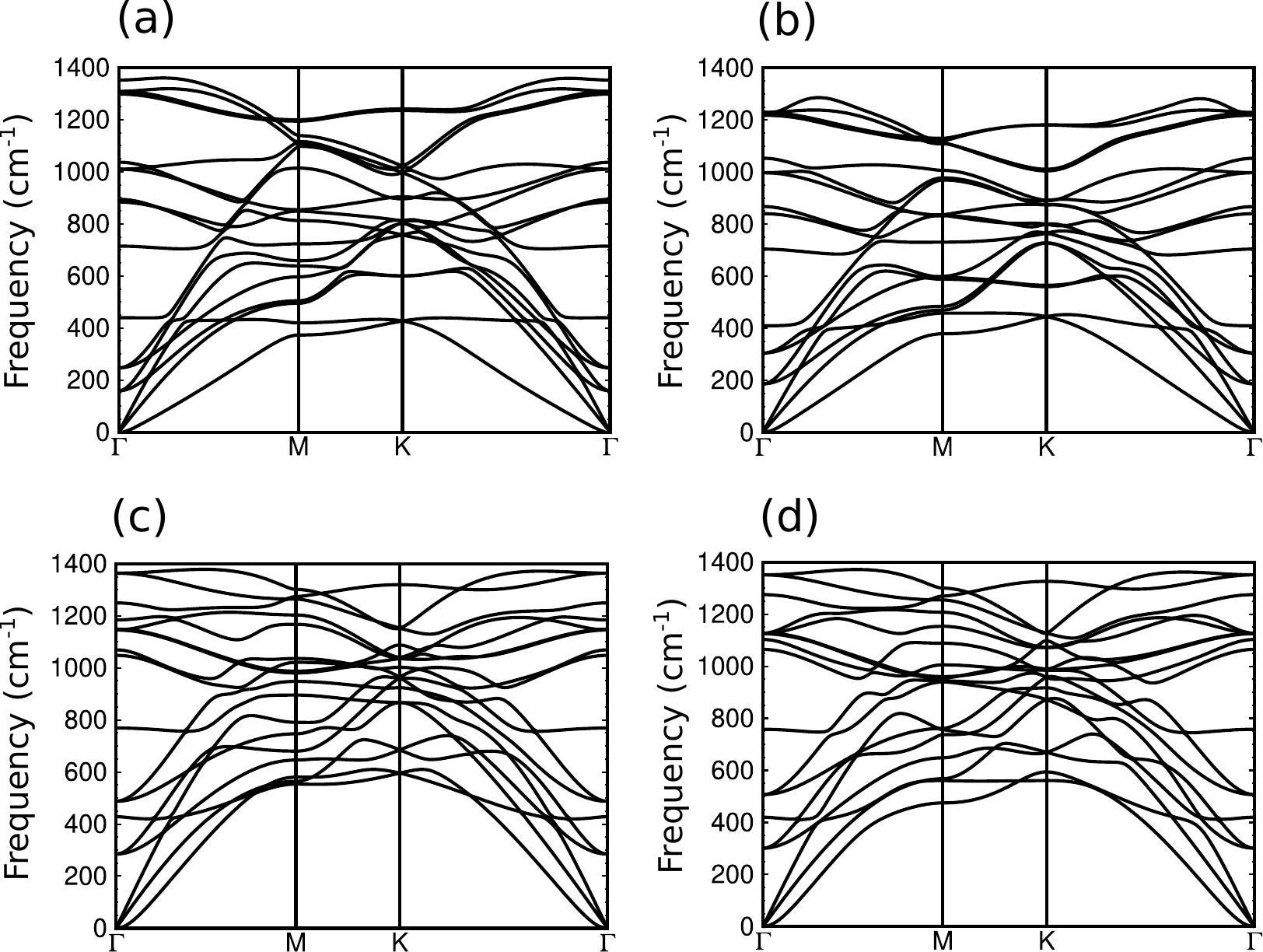}
\caption{Phonon dispersion branches of (a) AA$'$A$''$-C$_4$B$_2$, (b) ABC-C$_4$B$_2$, (c) AA$'$A$''$-C$_4$N$_2$, and (d) ABC-C$_4$N$_2$ along the main high-symmetry directions in the BZ of the hexagonal lattice.}
\label{phonon}
\end{figure}

Furthermore, the dynamic stability of the systems was investigated using the phonon theory, in which a system is considered stable when its vibrational spectrum contains only positive frequencies. The phonon dispersion curves of C$_4$B$_2$ and C$_4$N$_2$ compounds, in both AA$'$A$''$ and ABC stacking configurations, are presented in figure \ref{phonon}. All spectra show 18 phonon branches, related to the 6 atoms present in the primitive cell. All those systems are dynamically stable since there are only positive frequencies.  

\begin{figure*}[hbt]
\centering
\includegraphics[scale = 0.38, trim={0cm 0.0cm 0cm 0.0cm}, clip]{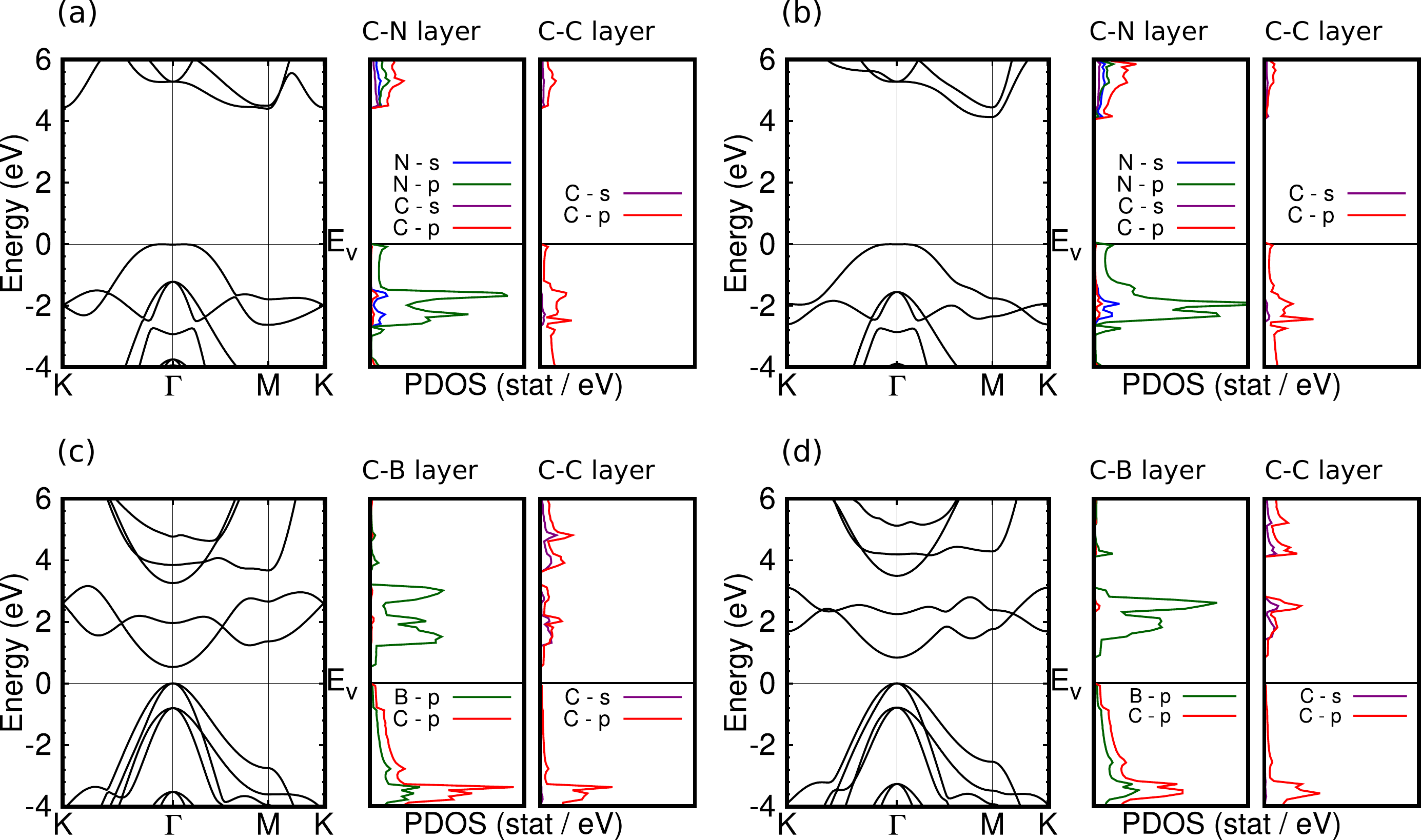}
\caption{Electronic band structures along the main high-symmetry directions of the BZ and PDOS, obtained with the optB88-vdW approach for the exchange-correlation energy: (a) AA$ ' $A$ '' $-C$_4$N$_2$, (b) ABC-C$_4$N$_2$, (c) AA$ ' $A$ '' $-C$_4$B$_2$, and (d) ABC-C$_4$B$_2$. The PDOS on the C and X $s$-orbitals are given in purple and blue, respectively, and on the C and X $p$-orbitals are given in red and green, respectively. E$_{\rm v}$ represents the VBM. Energies and PDPS are given in eV and states/eV, respectively.}
\label{band_pdos}
\end{figure*} 

Next, we computed the elastic constants of the systems using equation (\ref{eq_elastic_energy}) to verify their mechanical stability, according to the Born stability criteria ($C_{11}>0$ and $C_{12} < C_{11}$) \cite{mouhat2014necessary}. Table \ref{tab:table_elastic} presents the elastic constants $C_{11}$, $C_{12}$, and $C_{44} = \left(C_{11} - C_{12}\right)/2$, Young's modulus $\displaystyle Y^{\rm 2D} = {\left(C_{11}^2 - C_{12}^2\right)/C_{11}}$, and the Poisson ratio $\displaystyle \nu= {C_{12}/C_{11}}$ of C$_4$X$_2$ trilayers (X = B or N), as well as those of several other 2D materials for comparison.
\begin{table}[b!]
\label{tab:table_elastic}
\caption{Elastic constants $C_{11}$, $C_{12}$, and $C_{44}$, Young's modulus $Y^{\rm 2D}$, Poisson ratio $\sigma$, formal density $\rho_{\rm 2D}$, longitudinal $v_{\rm LA}$ and transverse $v_{\rm TA}$ acoustic velocities of C$_4$X$_2$ (X = B or N), graphene, and other 2D diamonds. Elastic constants and Young’s modulus are given in N/m, Poisson ratio is dimensionless, formal density and velocities are given in $10^{-7}$ kg/m$^2$ and km/s, respectively. The results with the $^{*}$ symbols were obtained using the data from the table and the equations described in this paper.
}
\vspace{0.15cm}
\begin{ruledtabular}
\begin{tabular}{lrccccccc}
\textrm{System} & \textrm{$\!\!C_{11}$ } & \textrm{$\!\!C_{12}$} & \textrm{$C_{44}$} & \textrm{$Y^{\rm 2D}$} & \textrm{$\sigma$} & $\rho_{\rm 2D}$ & $v_{\rm LA}$ & $v_{\rm TA}$ \\
& & & & & & & \\[-3.2mm]
\hline
& & & & & & & \\[-3mm]
AA$'$A$''$-C$_4$N$_2$ & $\!\!\!\!$816  & $\!\!$85 &  366 &  808  &  0.10 & 24.8 & 18.1 & 12.1 \\
ABC-C$_4$N$_2$ & $\!\!\!\!$777  & $\!\!$82 &  348 &  769 &   0.11 & 24.5 & 17.8 & 11.9  \\
& & & & & & & \\[-2.8mm]
AA$'$A$''$-C$_4$B$_2$ & $\!\!\!\!$627   & $\!\!$88 &  270  &  615 &  0.14 & 18.9 & 18.2 & 11.9  \\
ABC-C$_4$B$_2$ &  $\!\!\!\!$609   &$\!\!$92 &  259 &  595 &  0.15 & 18.7 & 18.0 & 11.7   \\
& & & & & & & \\[-2.8mm]
Graphene &  354\footnote{Reference \cite{cadelano2010elastic}}  & 60$^{\text{a}}$  & 147$^{*}$ & 340\footnote{Reference \cite{lee2008measurement} - experiments} & 0.18\footnote{Reference \cite{chernozatonskii2011influence}}  & 7.55$^{\text{c}}$ & 21.6$^{*}$ & 13.9$^{*}$ \\ 
HD &  474$^{\text{c}}$ & 36$^{\text{c}}$  & 219$^{*}$ & 471$^{*}$   & 0.08$^{\text{c}}$ & 14.9$^{\text{c}}$ & 17.8$^{\text{c}}$ & 12.2$^{\text{c}}$   \\
ABC-HD &  718$^{\text{c}}$  &58$^{\text{c}}$  & 330 & 713$^{*}$   & 0.08$^{\text{c}}$ & 22.2$^{\text{c}}$ & 18.0$^{\text{c}}$  & 12.2$^{\text{c}}$  \\
FD &  485\footnote{Reference \cite{cheng2020high}}   &49$^{*}$  & 218$^{*}$ & 480$^{\text{d}}$    & 0.10$^{*}$ & & 14.0\footnote{Reference \cite{sorokin2021two}} & 9.3$^{\text{e}}$   \\
NCCN &  568\footnote{Reference \cite{pakornchote2020roles}}  &66$^{\text{f}}$   & 243$^{\text{f}}$  & 560$^{*}$  & 0.12$^{*}$ & & & \\  
Diamond & 1079$^{\text{f}}$   &124$^{\text{f}}$  & 578$^{\text{f}}$  &   &  &  & 18.3$^{\text{c}}$  & 12.4$^{\text{c}}$  \\
\end{tabular}
\label{tab:table_elastic}
\end{ruledtabular}
\end{table}
 Accordingly, the C$_4$X$_2$ structures are mechanically stable since they satisfy the Born criteria, agreeing with the phonon dispersion spectra shown in figure \ref{phonon}. The C$_4$X$_2$ nanosheets present high Young's modulus values and characteristics of isotropic systems since their Poisson ratio $\sigma$ values are lower than 0.5 \cite{gercek2007poisson, hess2021bonding}. 

\newpage
Additionally, we estimated the longitudinal and the transversal acoustic velocities, given respectively by
\begin{equation}
v_{\rm LA} = \left( \frac{C_{11}}{\rho_{\rm 2D}}\right)^{1/2}\!\! \quad \mbox{and} \quad  
v_{\rm TA} = \left(\frac{C_{44}}{2 \rho_{\rm 2D}}\right)^{1/2}\!\!, 
\end{equation}
where $\rho_{\rm 2D}$ is the formal density, allowing comparison among systems, independent of their thickness \cite{sorokin2021two}. The velocity values, listed in Table \ref{tab:table_elastic}, suggest that the stiffness of the C$_4$X$_2$ systems is comparable with that of the diamond.

Following, we studied the electronic band structures and the projected density of states (PDOS) of the C$_4$X$_2$ systems, displayed in figure \ref{band_pdos} (a)-(d). The electronic band structures and PDOS of AA$'$A$''$-C$_4$N$_2$ and ABC-C$_4$N$_2$ nanosheets, presented in figures \ref{band_pdos} (a) and (b), exhibit some differences, despite their analogous PDOS, showing that the N $p$-orbitals dominate at valence band maximum (VBM) while the conduction band minimum (CBM) are mostly characterized by a mixture of $s$- and $p$-orbitals of N and C atoms. Both systems present the VBM around the $\Gamma$-point with a Mexican-hat dispersion, in which the two peaks lie on the $\Gamma$-K and $\Gamma$-M lines. The height of the Mexican-hat band at  $\Gamma$ point is 0.01 and 0.001 eV for AA$'$A$''$-C$_4$N$_2$ and ABC-C$_4$N$_2$, respectively. However, the CBM of ABC-C$_4$N$_2$ is well defined at the M-valley, while in AA$'$A$''$-C$_4$N$_2$ although the CBM is located at the M-valley,  the energy of the K-point is very close to the M-point one. On the other hand,  the AA$ ' $A$ '' $-C$_4$B$_2$ and ABC-C$_4$B$_2$ nanosheets have very similar band structures, as shown in figures \ref{band_pdos} (c) and (d). Both systems present direct band gaps, where the doubly degenerated VBM and the CBM are located at the $\Gamma$-point, in which the B $p$-orbitals dominate at the CBM and the VBM is described by a combination of B $p$-orbitals and C $p$-orbitals. 

\begin{table*}[hbt]
\begin{center}
\caption{Electronic band gap values of C$_4$X$_2$ (X = B or N) nanosheets (in eV), obtained with two different approximations for the exchange-correlation functional: optB88-vdW ($E_g^{\rm wdW}$) \cite{klimevs2009} and HSE ($E_g^{\rm HSE}$) \cite{heyd2003hybrid}. The PD band gap value is also displayed. The electron ($m^*_{e}$) and hole ($m^*_{h}$) effective masses, in $m_{0}$ units, obtained with the optB88-vdW approach are also shown. The VBM of C$_4$B$_2$ systems are doubly degenerated at $\Gamma$ and have two values for hole carrier. The CBM of C$_4$N$_2$ systems displays two effective masses around the M-valley, $m_e^{*M \rightarrow \Gamma}$ and $m_e^{*M \rightarrow K}$. } 
\begin{ruledtabular}
\begin{tabular}{lcccccccc}
\textrm{System} & $E_g^{\rm vdW}$ & $E_g^{\rm PBE}$  & $E_g^{\rm HSE}$ & Band gap & $m^*_{h}$ & $m^*_{e}$ & $m_e^{*M \rightarrow \Gamma}$ & $m_e^{*M \rightarrow K}$ \\ 
&  & \\[-3mm]
\hline
&  & \\[-2.8mm]
AA$'$A$''$-C$_4$N$_2$ & 4.40 &  & 5.56 & indirect & 6.23 &  & 2.94 &  0.30 \\
ABC-C$_4$N$_2$ & 4.13 &   & 5.42 & indirect  & 16.77 &  & 2.61 & 0.47  \\
&  & \\[-2.5mm]
AA$'$A$''$-C$_4$B$_2$ & 0.53 &  & 1.64 & direct &   0.34 (0.68) & 1.22  \\ 
ABC-C$_4$B$_2$ & 0.84  & & 1.97 & direct & 0.36 (0.80) & 1.30 \\ 
&  & \\[-2.3mm]
HD\footnote{Reference \cite{cheng2020high}}  & & 3.32  & & direct & 0.21 (0.58) & 1.11    \\
&  & \\[-2.3mm]
FD$^{\text{a}}$ & &4.04  & & direct & 0.37 (1.13) & 0.55    \\
&  & \\[-2.3mm]
PD\footnote{Reference \cite{ke2020synthesis}. The ($\overline{2}110$)-oriented h-diamane with 3 graphene layers.} &  &  & 2.70 & indirect & &\\   
\end{tabular}
\label{tab:table_electronic}
\end{ruledtabular}
\end{center}
\end{table*}

As discussed in the introduction, the 2D diamond systems present non-zero band gaps with characteristics that depend on several factors, such as doping with different functional groups. Herein, we are working with B and N atoms as X-doping elements,  which the B atom belongs to group-III elements of the periodic table, with a $2s^22p^1$ valence electronic configuration, and the N atom belongs to group-V elements,   with a $2s^22p^3$ valence electronic configuration. As a result, we found a wider indirect band gap for the C$_4$N$_2$ nanosheets, and a narrower direct band gap for the C$_4$B$_2$ systems when compared to PD. Table \ref{tab:table_electronic} displays the band gap values of the C$_4$X$_2$ nanosheets  obtained with the optB88-vdW \cite{klimevs2009} ($E_g^{\rm vdW}$) and the hybrid HSE  \cite{heyd2003hybrid} ($E_g^{\rm HSE}$) functional approaches for the exchange-correlation energy. For comparison, we also included the HD and FD  band gap values acquired with PBE functional \cite{cheng2020high} and of the three-layer graphene PD obtained with hybrid HSE functional \cite{ke2020synthesis}. The band gap width of the latter is 2.70 eV, while for the AA$ ' $A$ '' $-C$_4$N$_2$, ABC-C$_4$N$_2$, AA$ ' $A$ '' $-C$_4$B$_2$, and ABC-C$_4$B$_2$ functionalized compounds they are 5.56, 5.42, 1.64 eV, and 1.97 eV, respectively, using the same exchange-correlation functional. The C$_4$N$_2$ nanosheets have band gap width values, obtained with the optB88-vdW approach, similar to those of the HD and FD systems obtained with the PBE approximation.

Furthermore, since effective masses ($m^*$) can be used to investigate electronic transport under the influence of electric fields or carrier gradients, we estimated them by fitting parabolic functions to the CBM and VBM via the formula as follows:
\begin{equation}
    \frac{1}{m^*} = \frac{1}{\hbar^2} \Bigg| \frac{\partial^2 E(k)}{\partial k^2} \Bigg|
\end{equation}

\noindent
where $E(k)$ and $k$ are the energy and the wave vector of the CBM or VBM. The values of the effective masses depend on the curvature radius of the electronic band around the band-edge position, i.e., a larger curvature radius implies a heavier effective mass.

\begin{table*}[hbt]
\begin{center}
\caption{Summary of the qualitative properties and potential applications of C$_4$X$_2$ (X = B or N) systems. The column indicates dynamic (DS), mechanic (MS), and thermodynamic (TS) stability, Young's modulus $Y^{\rm 2D}$, and effective masses ($m^*$).}
\begin{ruledtabular}
\begin{tabular}{lccccccccc}
\textrm{System} & DS & MS & TS (300 K) & TS (1000 K) & \textrm{$Y^{\rm 2D}$} & Stiffness & Band gap & $m^*$ &  Potential applications   \\ 
&  & \\[-3mm]
\hline
&  & \\[-2.8mm]
C$_4$N$_2$ &  Yes & Yes & Yes & Yes  &high  &high  & indirect &heavier  &host material for   \\
          & & &  & & & & &  &single photon emission   \\
          & & & & & & & &  &quantum computing/biosensors   \\
\hline 
C$_4$B$_2$ & Yes & Yes & Yes & No  & high & high  &  direct & lighter  & nano-optics/nanoelectronics    \\ 
\end{tabular}
\label{tab:table_applications}
\end{ruledtabular}
\end{center}
\end{table*}

The electron ($m^*_e$) and hole ($m^*_h$) effective masses, in $m_0$ units, calculated with the optB88-vdW approach, are presented in table \ref{tab:table_electronic}. In both stacking structures of the C$_4$B$_2$ nanosheets, the effective masses of electrons and holes are comparable to the ones of HD and FD systems, which present extraordinary carriers mobility \cite{cheng2020high}. Furthermore, these estimated effective masses are similar to the $m_e^*= 1.06\,m_0$ and $m_h^*= 0.59\,m_0$ of silicon at a temperature of 4K \cite{green1990intrinsic}. Regarding C$_4$N$_2$ nanosheets, the $m^*_h$ is the magnitude of the effective mass at  $\Gamma$-point, i.e., we fitted a parabolic function considering the minimum located at the $\Gamma$-point \cite{wickramaratne2015electronic}. Accordingly, the $m^*_h$ depends on the height of the Mexican-hat band and the radius centered at $\Gamma$-point around band edges \cite{wickramaratne2015electronic}. On the other hand, the $m^*_e$ displays two effective masses around the M-valley. $m_e^{*M \rightarrow \Gamma}$ shows a high electron effective mass, much higher than $m_e^{*M \rightarrow K}$. $m_e^{*M \rightarrow \Gamma}$ is five times larger than $m_e^{*M \rightarrow K}$ for ABC-C$_4$N$_2$ and nine times larger for AA$ ' $A$ '' $-C$_4$N$_2$.

Considering the stable structures presented previously and their physical properties, it is interesting to explore their potential applications. Further investigations could explore the applicability of C$_4$N$_2$ and C$_4$B$_2$ structures as building blocks to build 2D/3D systems, such as van der Waals heterostructures, with different properties \cite{novoselov20162d,ipaves2019carbon, garcia2009}. Moreover, wide band gap materials, such as hexagonal boron nitride (h-BN), serve as a substrate for high-mobility 2D devices \cite{chaves2020bandgap},  a host material for single-photon emitter defect-centers for quantum computing and biosensors \cite{sorokin2021two}, etc. Therefore, the C$_4$N$_2$ nanosheets seem appropriate for these kinds of applications. Finally, the C$_4$B$_2$ nanosheets presented direct band gaps, in the 1.6 - 2.0 eV range, being more favorable for optoelectronics applications than C$_4$N$_2$ ones, which have indirect band gaps in the 5.4 - 5.6 eV range \cite{chaves2020bandgap}. In particular, the small effective masses and high elastic modulus of C$_4$B$_2$ systems may contribute to high electron mobility \cite{cheng2020high}, being suitable to be applied in photovoltaic cells. Table \ref{tab:table_applications} summarizes their properties and the respective potential applications.

In conclusion, we performed an {\it ab-initio} investigation on the structural, thermodynamic, dynamic, elastic, and electronic properties of C$_4$X$_2$ (X = B or N) systems. According to AIMD simulations, phonon calculations, and the Born stability criteria, all the nanosheets are thermodynamically, dynamically, and mechanically stable. Furthermore, the systems presented standard enthalpy of formation close to the recently synthesized pristine 2D diamond composed of three graphene layers. Elastic properties indicated that those nanosheets possess a high Young's modulus values and characteristics of isotropic systems, and the estimated longitudinal and transversal acoustic velocities revealed that their stiffness is comparable with that of the diamond. Finally, the systems' electronic properties presented some differences, in which C$_4$N$_2$ structures exhibited wide indirect band gaps and heavier effective masses, while the C$_4$B$_2$ ones had narrow direct band gaps and lighter effective masses. These results provide chemical routes to tune the electronic properties of 2D diamonds by doping them for specific applications, such as optoelectronic devices.

\begin{acknowledgments}

Brazilian Federal Government Agencies CAPES (Grants 88882.332907/2019-01 and 88887.371193/2019-00), CNPq (Grants 314884/2021-1, 302800/2022-0, and 150595/2023-9) and FAPESP (Grant 22/10095-8) partially supported this investigation. The authors acknowledge the National Laboratory for Scientific Computing (\href{http://sdumont.lncc.br}{LNCC/MCTI}, Brazil) for providing HPC resources of the Santos Dumont supercomputer, Centro Nacional de Processamento de Alto Desempenho em São Paulo (\href{https://www.cenapad.unicamp.br/}{CENAPAD-SP}, Brazil) and SNIC-UPPMAX, SNIC-HPC2N, and SNIC-NSC centers under the Swedish National Infrastructure for Computing (SNIC) resources.

\end{acknowledgments}
 
 \newpage
\bibliography{apssamp}

\providecommand{\noopsort}[1]{}\providecommand{\singleletter}[1]{#1}%
\begin{thebibliography}{46}%
\makeatletter
\providecommand \@ifxundefined [1]{%
 \@ifx{#1\undefined}
}%
\providecommand \@ifnum [1]{%
 \ifnum #1\expandafter \@firstoftwo
 \else \expandafter \@secondoftwo
 \fi
}%
\providecommand \@ifx [1]{%
 \ifx #1\expandafter \@firstoftwo
 \else \expandafter \@secondoftwo
 \fi
}%
\providecommand \natexlab [1]{#1}%
\providecommand \enquote  [1]{``#1''}%
\providecommand \bibnamefont  [1]{#1}%
\providecommand \bibfnamefont [1]{#1}%
\providecommand \citenamefont [1]{#1}%
\providecommand \href@noop [0]{\@secondoftwo}%
\providecommand \href [0]{\begingroup \@sanitize@url \@href}%
\providecommand \@href[1]{\@@startlink{#1}\@@href}%
\providecommand \@@href[1]{\endgroup#1\@@endlink}%
\providecommand \@sanitize@url [0]{\catcode `\\12\catcode `\$12\catcode
  `\&12\catcode `\#12\catcode `\^12\catcode `\_12\catcode `\%12\relax}%
\providecommand \@@startlink[1]{}%
\providecommand \@@endlink[0]{}%
\providecommand \url  [0]{\begingroup\@sanitize@url \@url }%
\providecommand \@url [1]{\endgroup\@href {#1}{\urlprefix }}%
\providecommand \urlprefix  [0]{URL }%
\providecommand \Eprint [0]{\href }%
\providecommand \doibase [0]{https://doi.org/}%
\providecommand \selectlanguage [0]{\@gobble}%
\providecommand \bibinfo  [0]{\@secondoftwo}%
\providecommand \bibfield  [0]{\@secondoftwo}%
\providecommand \translation [1]{[#1]}%
\providecommand \BibitemOpen [0]{}%
\providecommand \bibitemStop [0]{}%
\providecommand \bibitemNoStop [0]{.\EOS\space}%
\providecommand \EOS [0]{\spacefactor3000\relax}%
\providecommand \BibitemShut  [1]{\csname bibitem#1\endcsname}%
\let\auto@bib@innerbib\@empty
\bibitem [{\citenamefont {Novoselov}\ \emph {et~al.}(2004)\citenamefont
  {Novoselov}, \citenamefont {Geim}, \citenamefont {Morozov}, \citenamefont
  {Jiang}, \citenamefont {Zhang}, \citenamefont {Dubonos}, \citenamefont
  {Grigorieva},\ and\ \citenamefont {Firsov}}]{novoselov2004}%
  \BibitemOpen
  \bibfield  {author} {\bibinfo {author} {\bibfnamefont {K.~S.}\ \bibnamefont
  {Novoselov}}, \bibinfo {author} {\bibfnamefont {A.~K.}\ \bibnamefont {Geim}},
  \bibinfo {author} {\bibfnamefont {S.~V.}\ \bibnamefont {Morozov}}, \bibinfo
  {author} {\bibfnamefont {D.}~\bibnamefont {Jiang}}, \bibinfo {author}
  {\bibfnamefont {Y.}~\bibnamefont {Zhang}}, \bibinfo {author} {\bibfnamefont
  {S.~V.}\ \bibnamefont {Dubonos}}, \bibinfo {author} {\bibfnamefont {I.~V.}\
  \bibnamefont {Grigorieva}},\ and\ \bibinfo {author} {\bibfnamefont {A.~A.}\
  \bibnamefont {Firsov}},\ }\bibfield  {title} {\bibinfo {title} {Electric
  field effect in atomically thin carbon films},\ }\href@noop {} {\bibfield
  {journal} {\bibinfo  {journal} {Science}\ }\textbf {\bibinfo {volume}
  {306}},\ \bibinfo {pages} {666} (\bibinfo {year} {2004})}\BibitemShut
  {NoStop}%
\bibitem [{\citenamefont {Geim}\ and\ \citenamefont
  {Novoselov}(2007)}]{geim2007rise}%
  \BibitemOpen
  \bibfield  {author} {\bibinfo {author} {\bibfnamefont {A.~K.}\ \bibnamefont
  {Geim}}\ and\ \bibinfo {author} {\bibfnamefont {K.~S.}\ \bibnamefont
  {Novoselov}},\ }\bibfield  {title} {\bibinfo {title} {The rise of graphene},\
  }\href@noop {} {\bibfield  {journal} {\bibinfo  {journal} {Nature Materials}\
  }\textbf {\bibinfo {volume} {6}},\ \bibinfo {pages} {183} (\bibinfo {year}
  {2007})}\BibitemShut {NoStop}%
\bibitem [{\citenamefont {Gao}\ \emph {et~al.}(2018)\citenamefont {Gao},
  \citenamefont {Cao}, \citenamefont {Cellini}, \citenamefont {Berger},
  \citenamefont {De~Heer}, \citenamefont {Tosatti}, \citenamefont {Riedo},\
  and\ \citenamefont {Bongiorno}}]{gao2018ultrahard}%
  \BibitemOpen
  \bibfield  {author} {\bibinfo {author} {\bibfnamefont {Y.}~\bibnamefont
  {Gao}}, \bibinfo {author} {\bibfnamefont {T.}~\bibnamefont {Cao}}, \bibinfo
  {author} {\bibfnamefont {F.}~\bibnamefont {Cellini}}, \bibinfo {author}
  {\bibfnamefont {C.}~\bibnamefont {Berger}}, \bibinfo {author} {\bibfnamefont
  {W.~A.}\ \bibnamefont {De~Heer}}, \bibinfo {author} {\bibfnamefont
  {E.}~\bibnamefont {Tosatti}}, \bibinfo {author} {\bibfnamefont
  {E.}~\bibnamefont {Riedo}},\ and\ \bibinfo {author} {\bibfnamefont
  {A.}~\bibnamefont {Bongiorno}},\ }\bibfield  {title} {\bibinfo {title}
  {Ultrahard carbon film from epitaxial two-layer graphene},\ }\href@noop {}
  {\bibfield  {journal} {\bibinfo  {journal} {Nature Nanotechnology}\ }\textbf
  {\bibinfo {volume} {13}},\ \bibinfo {pages} {133} (\bibinfo {year}
  {2018})}\BibitemShut {NoStop}%
\bibitem [{\citenamefont {Sorokin}\ and\ \citenamefont
  {Yakobson}(2021)}]{sorokin2021two}%
  \BibitemOpen
  \bibfield  {author} {\bibinfo {author} {\bibfnamefont {P.~B.}\ \bibnamefont
  {Sorokin}}\ and\ \bibinfo {author} {\bibfnamefont {B.~I.}\ \bibnamefont
  {Yakobson}},\ }\bibfield  {title} {\bibinfo {title} {Two-dimensional
  diamond—diamane: current state and further prospects},\ }\href@noop {}
  {\bibfield  {journal} {\bibinfo  {journal} {Nano Letters}\ }\textbf {\bibinfo
  {volume} {21}},\ \bibinfo {pages} {5475} (\bibinfo {year}
  {2021})}\BibitemShut {NoStop}%
\bibitem [{\citenamefont {Qin}\ \emph {et~al.}(2021)\citenamefont {Qin},
  \citenamefont {Wu},\ and\ \citenamefont {Gou}}]{qin2021diamane}%
  \BibitemOpen
  \bibfield  {author} {\bibinfo {author} {\bibfnamefont {G.}~\bibnamefont
  {Qin}}, \bibinfo {author} {\bibfnamefont {L.}~\bibnamefont {Wu}},\ and\
  \bibinfo {author} {\bibfnamefont {H.}~\bibnamefont {Gou}},\ }\bibfield
  {title} {\bibinfo {title} {Diamane: design, synthesis, properties, and
  challenges},\ }\href@noop {} {\bibfield  {journal} {\bibinfo  {journal}
  {Functional Diamond}\ }\textbf {\bibinfo {volume} {1}},\ \bibinfo {pages}
  {83} (\bibinfo {year} {2021})}\BibitemShut {NoStop}%
\bibitem [{\citenamefont {Piazza}\ \emph {et~al.}(2019)\citenamefont {Piazza},
  \citenamefont {Gough}, \citenamefont {Monthioux}, \citenamefont {Puech},
  \citenamefont {Gerber}, \citenamefont {Wiens}, \citenamefont {Paredes},\ and\
  \citenamefont {Ozoria}}]{piazza2019low}%
  \BibitemOpen
  \bibfield  {author} {\bibinfo {author} {\bibfnamefont {F.}~\bibnamefont
  {Piazza}}, \bibinfo {author} {\bibfnamefont {K.}~\bibnamefont {Gough}},
  \bibinfo {author} {\bibfnamefont {M.}~\bibnamefont {Monthioux}}, \bibinfo
  {author} {\bibfnamefont {P.}~\bibnamefont {Puech}}, \bibinfo {author}
  {\bibfnamefont {I.}~\bibnamefont {Gerber}}, \bibinfo {author} {\bibfnamefont
  {R.}~\bibnamefont {Wiens}}, \bibinfo {author} {\bibfnamefont
  {G.}~\bibnamefont {Paredes}},\ and\ \bibinfo {author} {\bibfnamefont
  {C.}~\bibnamefont {Ozoria}},\ }\bibfield  {title} {\bibinfo {title} {Low
  temperature, pressureless sp$^2$ to sp$^3$ transformation of ultrathin,
  crystalline carbon films},\ }\href@noop {} {\bibfield  {journal} {\bibinfo
  {journal} {Carbon}\ }\textbf {\bibinfo {volume} {145}},\ \bibinfo {pages}
  {10} (\bibinfo {year} {2019})}\BibitemShut {NoStop}%
\bibitem [{\citenamefont {Grayfer}\ \emph {et~al.}(2013)\citenamefont
  {Grayfer}, \citenamefont {Makotchenko}, \citenamefont {Kibis}, \citenamefont
  {Boronin}, \citenamefont {Pazhetnov}, \citenamefont {Zaikovskii},\ and\
  \citenamefont {Fedorov}}]{grayfer2013synthesis}%
  \BibitemOpen
  \bibfield  {author} {\bibinfo {author} {\bibfnamefont {E.~D.}\ \bibnamefont
  {Grayfer}}, \bibinfo {author} {\bibfnamefont {V.~G.}\ \bibnamefont
  {Makotchenko}}, \bibinfo {author} {\bibfnamefont {L.~S.}\ \bibnamefont
  {Kibis}}, \bibinfo {author} {\bibfnamefont {A.~I.}\ \bibnamefont {Boronin}},
  \bibinfo {author} {\bibfnamefont {E.~M.}\ \bibnamefont {Pazhetnov}}, \bibinfo
  {author} {\bibfnamefont {V.~I.}\ \bibnamefont {Zaikovskii}},\ and\ \bibinfo
  {author} {\bibfnamefont {V.~E.}\ \bibnamefont {Fedorov}},\ }\bibfield
  {title} {\bibinfo {title} {Synthesis, properties, and dispersion of few-layer
  graphene fluoride},\ }\href@noop {} {\bibfield  {journal} {\bibinfo
  {journal} {Chemistry--An Asian Journal}\ }\textbf {\bibinfo {volume} {8}},\
  \bibinfo {pages} {2015} (\bibinfo {year} {2013})}\BibitemShut {NoStop}%
\bibitem [{\citenamefont {Ke}\ \emph {et~al.}(2020)\citenamefont {Ke},
  \citenamefont {Zhang}, \citenamefont {Chen}, \citenamefont {Yin},
  \citenamefont {Wang}, \citenamefont {Tzeng}, \citenamefont {Lin},
  \citenamefont {Dong}, \citenamefont {Liu}, \citenamefont {Tse} \emph
  {et~al.}}]{ke2020synthesis}%
  \BibitemOpen
  \bibfield  {author} {\bibinfo {author} {\bibfnamefont {F.}~\bibnamefont
  {Ke}}, \bibinfo {author} {\bibfnamefont {L.}~\bibnamefont {Zhang}}, \bibinfo
  {author} {\bibfnamefont {Y.}~\bibnamefont {Chen}}, \bibinfo {author}
  {\bibfnamefont {K.}~\bibnamefont {Yin}}, \bibinfo {author} {\bibfnamefont
  {C.}~\bibnamefont {Wang}}, \bibinfo {author} {\bibfnamefont {Y.-K.}\
  \bibnamefont {Tzeng}}, \bibinfo {author} {\bibfnamefont {Y.}~\bibnamefont
  {Lin}}, \bibinfo {author} {\bibfnamefont {H.}~\bibnamefont {Dong}}, \bibinfo
  {author} {\bibfnamefont {Z.}~\bibnamefont {Liu}}, \bibinfo {author}
  {\bibfnamefont {J.~S.}\ \bibnamefont {Tse}}, \emph {et~al.},\ }\bibfield
  {title} {\bibinfo {title} {Synthesis of atomically thin hexagonal diamond
  with compression},\ }\href@noop {} {\bibfield  {journal} {\bibinfo  {journal}
  {Nano Letters}\ }\textbf {\bibinfo {volume} {20}},\ \bibinfo {pages} {5916}
  (\bibinfo {year} {2020})}\BibitemShut {NoStop}%
\bibitem [{\citenamefont {Pakornchote}\ \emph {et~al.}(2020)\citenamefont
  {Pakornchote}, \citenamefont {Ektarawong}, \citenamefont {Busayaporn},
  \citenamefont {Pinsook},\ and\ \citenamefont
  {Bovornratanaraks}}]{pakornchote2020roles}%
  \BibitemOpen
  \bibfield  {author} {\bibinfo {author} {\bibfnamefont {T.}~\bibnamefont
  {Pakornchote}}, \bibinfo {author} {\bibfnamefont {A.}~\bibnamefont
  {Ektarawong}}, \bibinfo {author} {\bibfnamefont {W.}~\bibnamefont
  {Busayaporn}}, \bibinfo {author} {\bibfnamefont {U.}~\bibnamefont
  {Pinsook}},\ and\ \bibinfo {author} {\bibfnamefont {T.}~\bibnamefont
  {Bovornratanaraks}},\ }\bibfield  {title} {\bibinfo {title} {Roles of
  nitrogen substitution and surface reconstruction in stabilizing nonpassivated
  single-layer diamond},\ }\href@noop {} {\bibfield  {journal} {\bibinfo
  {journal} {Physical Review B}\ }\textbf {\bibinfo {volume} {102}},\ \bibinfo
  {pages} {075418} (\bibinfo {year} {2020})}\BibitemShut {NoStop}%
\bibitem [{\citenamefont {Ipaves}\ \emph {et~al.}(2019)\citenamefont {Ipaves},
  \citenamefont {Justo},\ and\ \citenamefont {Assali}}]{ipaves2019carbon}%
  \BibitemOpen
  \bibfield  {author} {\bibinfo {author} {\bibfnamefont {B.}~\bibnamefont
  {Ipaves}}, \bibinfo {author} {\bibfnamefont {J.~F.}\ \bibnamefont {Justo}},\
  and\ \bibinfo {author} {\bibfnamefont {L.~V.~C.}\ \bibnamefont {Assali}},\
  }\bibfield  {title} {\bibinfo {title} {Carbon-related bilayers: Nanoscale
  building blocks for self-assembly nanomanufacturing},\ }\href@noop {}
  {\bibfield  {journal} {\bibinfo  {journal} {The Journal of Physical Chemistry
  C}\ }\textbf {\bibinfo {volume} {123}},\ \bibinfo {pages} {23195} (\bibinfo
  {year} {2019})}\BibitemShut {NoStop}%
\bibitem [{\citenamefont {He}\ \emph {et~al.}(2020)\citenamefont {He},
  \citenamefont {Zhang}, \citenamefont {Li},\ and\ \citenamefont
  {Zhang}}]{he2020electric}%
  \BibitemOpen
  \bibfield  {author} {\bibinfo {author} {\bibfnamefont {C.}~\bibnamefont
  {He}}, \bibinfo {author} {\bibfnamefont {M.}~\bibnamefont {Zhang}}, \bibinfo
  {author} {\bibfnamefont {T.}~\bibnamefont {Li}},\ and\ \bibinfo {author}
  {\bibfnamefont {W.}~\bibnamefont {Zhang}},\ }\bibfield  {title} {\bibinfo
  {title} {Electric field-modulated high sensitivity and selectivity for nh$_3$
  on $\alpha$-c$_2$n$_2$ nanosheet: Insights from dft calculations},\
  }\href@noop {} {\bibfield  {journal} {\bibinfo  {journal} {Applied Surface
  Science}\ }\textbf {\bibinfo {volume} {505}},\ \bibinfo {pages} {144619}
  (\bibinfo {year} {2020})}\BibitemShut {NoStop}%
\bibitem [{\citenamefont {Tan}\ \emph {et~al.}(2021)\citenamefont {Tan},
  \citenamefont {Nie}, \citenamefont {Ao}, \citenamefont {Sun}, \citenamefont
  {An},\ and\ \citenamefont {Wang}}]{tan2020novel}%
  \BibitemOpen
  \bibfield  {author} {\bibinfo {author} {\bibfnamefont {L.}~\bibnamefont
  {Tan}}, \bibinfo {author} {\bibfnamefont {C.}~\bibnamefont {Nie}}, \bibinfo
  {author} {\bibfnamefont {Z.}~\bibnamefont {Ao}}, \bibinfo {author}
  {\bibfnamefont {H.}~\bibnamefont {Sun}}, \bibinfo {author} {\bibfnamefont
  {T.}~\bibnamefont {An}},\ and\ \bibinfo {author} {\bibfnamefont
  {S.}~\bibnamefont {Wang}},\ }\bibfield  {title} {\bibinfo {title} {Novel
  two-dimensional crystalline carbon nitrides beyond gc$_3$n$_4$: structure and
  applications},\ }\href@noop {} {\bibfield  {journal} {\bibinfo  {journal}
  {Journal of Materials Chemistry A}\ }\textbf {\bibinfo {volume} {9}},\
  \bibinfo {pages} {17} (\bibinfo {year} {2021})}\BibitemShut {NoStop}%
\bibitem [{\citenamefont {Zhu}\ and\ \citenamefont
  {Zhang}(2019)}]{zhu2019suppressed}%
  \BibitemOpen
  \bibfield  {author} {\bibinfo {author} {\bibfnamefont {L.}~\bibnamefont
  {Zhu}}\ and\ \bibinfo {author} {\bibfnamefont {T.}~\bibnamefont {Zhang}},\
  }\bibfield  {title} {\bibinfo {title} {Suppressed thermal conductivity in
  fluorinated diamane: optical phonon dominant thermal transport},\ }\href@noop
  {} {\bibfield  {journal} {\bibinfo  {journal} {Applied Physics Letters}\
  }\textbf {\bibinfo {volume} {115}},\ \bibinfo {pages} {151904} (\bibinfo
  {year} {2019})}\BibitemShut {NoStop}%
\bibitem [{\citenamefont {Cellini}\ \emph {et~al.}(2018)\citenamefont
  {Cellini}, \citenamefont {Lavini}, \citenamefont {Cao}, \citenamefont
  {de~Heer}, \citenamefont {Berger}, \citenamefont {Bongiorno},\ and\
  \citenamefont {Riedo}}]{cellini2018epitaxial}%
  \BibitemOpen
  \bibfield  {author} {\bibinfo {author} {\bibfnamefont {F.}~\bibnamefont
  {Cellini}}, \bibinfo {author} {\bibfnamefont {F.}~\bibnamefont {Lavini}},
  \bibinfo {author} {\bibfnamefont {T.}~\bibnamefont {Cao}}, \bibinfo {author}
  {\bibfnamefont {W.}~\bibnamefont {de~Heer}}, \bibinfo {author} {\bibfnamefont
  {C.}~\bibnamefont {Berger}}, \bibinfo {author} {\bibfnamefont
  {A.}~\bibnamefont {Bongiorno}},\ and\ \bibinfo {author} {\bibfnamefont
  {E.}~\bibnamefont {Riedo}},\ }\bibfield  {title} {\bibinfo {title} {Epitaxial
  two-layer graphene under pressure: diamene stiffer than diamond},\
  }\href@noop {} {\bibfield  {journal} {\bibinfo  {journal} {FlatChem}\
  }\textbf {\bibinfo {volume} {10}},\ \bibinfo {pages} {8} (\bibinfo {year}
  {2018})}\BibitemShut {NoStop}%
\bibitem [{\citenamefont {Cheng}\ \emph {et~al.}(2020)\citenamefont {Cheng},
  \citenamefont {Liu},\ and\ \citenamefont {Liu}}]{cheng2020high}%
  \BibitemOpen
  \bibfield  {author} {\bibinfo {author} {\bibfnamefont {T.}~\bibnamefont
  {Cheng}}, \bibinfo {author} {\bibfnamefont {Z.}~\bibnamefont {Liu}},\ and\
  \bibinfo {author} {\bibfnamefont {Z.}~\bibnamefont {Liu}},\ }\bibfield
  {title} {\bibinfo {title} {High elastic moduli, controllable bandgap and
  extraordinary carrier mobility in single-layer diamond},\ }\href@noop {}
  {\bibfield  {journal} {\bibinfo  {journal} {Journal of Materials Chemistry
  C}\ }\textbf {\bibinfo {volume} {8}},\ \bibinfo {pages} {13819} (\bibinfo
  {year} {2020})}\BibitemShut {NoStop}%
\bibitem [{\citenamefont {Kohn}\ and\ \citenamefont
  {Sham}(1965)}]{kohn1965self}%
  \BibitemOpen
  \bibfield  {author} {\bibinfo {author} {\bibfnamefont {W.}~\bibnamefont
  {Kohn}}\ and\ \bibinfo {author} {\bibfnamefont {L.~J.}\ \bibnamefont
  {Sham}},\ }\bibfield  {title} {\bibinfo {title} {Self-consistent equations
  including exchange and correlation effects},\ }\href@noop {} {\bibfield
  {journal} {\bibinfo  {journal} {Physical Review}\ }\textbf {\bibinfo {volume}
  {140}},\ \bibinfo {pages} {A1133} (\bibinfo {year} {1965})}\BibitemShut
  {NoStop}%
\bibitem [{\citenamefont {Kresse}\ and\ \citenamefont
  {Joubert}(1999)}]{kresse1999ultrasoft}%
  \BibitemOpen
  \bibfield  {author} {\bibinfo {author} {\bibfnamefont {G.}~\bibnamefont
  {Kresse}}\ and\ \bibinfo {author} {\bibfnamefont {D.}~\bibnamefont
  {Joubert}},\ }\bibfield  {title} {\bibinfo {title} {From ultrasoft
  pseudopotentials to the projector augmented-wave method},\ }\href@noop {}
  {\bibfield  {journal} {\bibinfo  {journal} {Physical Review B}\ }\textbf
  {\bibinfo {volume} {59}},\ \bibinfo {pages} {1758} (\bibinfo {year}
  {1999})}\BibitemShut {NoStop}%
\bibitem [{\citenamefont {Giannozzi}\ \emph {et~al.}(2009)\citenamefont
  {Giannozzi}, \citenamefont {Baroni}, \citenamefont {Bonini}, \citenamefont
  {Calandra}, \citenamefont {Car}, \citenamefont {Cavazzoni}, \citenamefont
  {Ceresoli}, \citenamefont {Chiarotti}, \citenamefont {Cococcioni},
  \citenamefont {Dabo} \emph {et~al.}}]{giannozzi2009quantum}%
  \BibitemOpen
  \bibfield  {author} {\bibinfo {author} {\bibfnamefont {P.}~\bibnamefont
  {Giannozzi}}, \bibinfo {author} {\bibfnamefont {S.}~\bibnamefont {Baroni}},
  \bibinfo {author} {\bibfnamefont {N.}~\bibnamefont {Bonini}}, \bibinfo
  {author} {\bibfnamefont {M.}~\bibnamefont {Calandra}}, \bibinfo {author}
  {\bibfnamefont {R.}~\bibnamefont {Car}}, \bibinfo {author} {\bibfnamefont
  {C.}~\bibnamefont {Cavazzoni}}, \bibinfo {author} {\bibfnamefont
  {D.}~\bibnamefont {Ceresoli}}, \bibinfo {author} {\bibfnamefont {G.~L.}\
  \bibnamefont {Chiarotti}}, \bibinfo {author} {\bibfnamefont {M.}~\bibnamefont
  {Cococcioni}}, \bibinfo {author} {\bibfnamefont {I.}~\bibnamefont {Dabo}},
  \emph {et~al.},\ }\bibfield  {title} {\bibinfo {title} {Quantum espresso: a
  modular and open-source software project for quantum simulations of
  materials},\ }\href@noop {} {\bibfield  {journal} {\bibinfo  {journal}
  {Journal of Physics: Condensed Matter}\ }\textbf {\bibinfo {volume} {21}},\
  \bibinfo {pages} {395502} (\bibinfo {year} {2009})}\BibitemShut {NoStop}%
\bibitem [{\citenamefont {Giannozzi}\ \emph {et~al.}(2017)\citenamefont
  {Giannozzi}, \citenamefont {Andreussi}, \citenamefont {Brumme}, \citenamefont
  {Bunau}, \citenamefont {Nardelli}, \citenamefont {Calandra}, \citenamefont
  {Car}, \citenamefont {Cavazzoni}, \citenamefont {Ceresoli},\ and\
  \citenamefont {Cococcioni}}]{giannozzi2017advanced}%
  \BibitemOpen
  \bibfield  {author} {\bibinfo {author} {\bibfnamefont {P.}~\bibnamefont
  {Giannozzi}}, \bibinfo {author} {\bibfnamefont {O.}~\bibnamefont
  {Andreussi}}, \bibinfo {author} {\bibfnamefont {T.}~\bibnamefont {Brumme}},
  \bibinfo {author} {\bibfnamefont {O.}~\bibnamefont {Bunau}}, \bibinfo
  {author} {\bibfnamefont {M.~B.}\ \bibnamefont {Nardelli}}, \bibinfo {author}
  {\bibfnamefont {M.}~\bibnamefont {Calandra}}, \bibinfo {author}
  {\bibfnamefont {R.}~\bibnamefont {Car}}, \bibinfo {author} {\bibfnamefont
  {C.}~\bibnamefont {Cavazzoni}}, \bibinfo {author} {\bibfnamefont
  {D.}~\bibnamefont {Ceresoli}},\ and\ \bibinfo {author} {\bibfnamefont
  {M.~{\it et al}.}\ \bibnamefont {Cococcioni}},\ }\bibfield  {title} {\bibinfo
  {title} {Advanced capabilities for materials modelling with quantum
  espresso},\ }\href@noop {} {\bibfield  {journal} {\bibinfo  {journal}
  {Journal of Physics: Condensed Matter}\ }\textbf {\bibinfo {volume} {29}},\
  \bibinfo {pages} {465901} (\bibinfo {year} {2017})}\BibitemShut {NoStop}%
\bibitem [{\citenamefont {Perdew}\ \emph {et~al.}(1996)\citenamefont {Perdew},
  \citenamefont {Burke},\ and\ \citenamefont
  {Ernzerhof}}]{perdew1996generalized}%
  \BibitemOpen
  \bibfield  {author} {\bibinfo {author} {\bibfnamefont {J.~P.}\ \bibnamefont
  {Perdew}}, \bibinfo {author} {\bibfnamefont {K.}~\bibnamefont {Burke}},\ and\
  \bibinfo {author} {\bibfnamefont {M.}~\bibnamefont {Ernzerhof}},\ }\bibfield
  {title} {\bibinfo {title} {Generalized gradient approximation made simple},\
  }\href@noop {} {\bibfield  {journal} {\bibinfo  {journal} {Physical Review
  Letters}\ }\textbf {\bibinfo {volume} {77}},\ \bibinfo {pages} {3865}
  (\bibinfo {year} {1996})}\BibitemShut {NoStop}%
\bibitem [{\citenamefont {Dion}\ \emph {et~al.}(2004)\citenamefont {Dion},
  \citenamefont {Rydberg}, \citenamefont {Schr{\"o}der}, \citenamefont
  {Langreth},\ and\ \citenamefont {Lundqvist}}]{dion2004}%
  \BibitemOpen
  \bibfield  {author} {\bibinfo {author} {\bibfnamefont {M.}~\bibnamefont
  {Dion}}, \bibinfo {author} {\bibfnamefont {H.}~\bibnamefont {Rydberg}},
  \bibinfo {author} {\bibfnamefont {E.}~\bibnamefont {Schr{\"o}der}}, \bibinfo
  {author} {\bibfnamefont {D.~C.}\ \bibnamefont {Langreth}},\ and\ \bibinfo
  {author} {\bibfnamefont {B.~I.}\ \bibnamefont {Lundqvist}},\ }\bibfield
  {title} {\bibinfo {title} {Van der waals density functional for general
  geometries},\ }\href@noop {} {\bibfield  {journal} {\bibinfo  {journal}
  {Physical Review Letters}\ }\textbf {\bibinfo {volume} {92}},\ \bibinfo
  {pages} {246401} (\bibinfo {year} {2004})}\BibitemShut {NoStop}%
\bibitem [{\citenamefont {Klime{\v{s}}}\ \emph {et~al.}(2009)\citenamefont
  {Klime{\v{s}}}, \citenamefont {Bowler},\ and\ \citenamefont
  {Michaelides}}]{klimevs2009}%
  \BibitemOpen
  \bibfield  {author} {\bibinfo {author} {\bibfnamefont {J.}~\bibnamefont
  {Klime{\v{s}}}}, \bibinfo {author} {\bibfnamefont {D.~R.}\ \bibnamefont
  {Bowler}},\ and\ \bibinfo {author} {\bibfnamefont {A.}~\bibnamefont
  {Michaelides}},\ }\bibfield  {title} {\bibinfo {title} {Chemical accuracy for
  the van der waals density functional},\ }\href@noop {} {\bibfield  {journal}
  {\bibinfo  {journal} {Journal of Physics: Condensed Matter}\ }\textbf
  {\bibinfo {volume} {22}},\ \bibinfo {pages} {022201} (\bibinfo {year}
  {2009})}\BibitemShut {NoStop}%
\bibitem [{\citenamefont {Heyd}\ \emph {et~al.}(2003)\citenamefont {Heyd},
  \citenamefont {Scuseria},\ and\ \citenamefont {Ernzerhof}}]{heyd2003hybrid}%
  \BibitemOpen
  \bibfield  {author} {\bibinfo {author} {\bibfnamefont {J.}~\bibnamefont
  {Heyd}}, \bibinfo {author} {\bibfnamefont {G.~E.}\ \bibnamefont {Scuseria}},\
  and\ \bibinfo {author} {\bibfnamefont {M.}~\bibnamefont {Ernzerhof}},\
  }\bibfield  {title} {\bibinfo {title} {Hybrid functionals based on a screened
  coulomb potential},\ }\href@noop {} {\bibfield  {journal} {\bibinfo
  {journal} {The Journal of Chemical Physics}\ }\textbf {\bibinfo {volume}
  {118}},\ \bibinfo {pages} {8207} (\bibinfo {year} {2003})}\BibitemShut
  {NoStop}%
\bibitem [{\citenamefont {Monkhorst}\ and\ \citenamefont
  {Pack}(1976)}]{monkhorst1976special}%
  \BibitemOpen
  \bibfield  {author} {\bibinfo {author} {\bibfnamefont {H.~J.}\ \bibnamefont
  {Monkhorst}}\ and\ \bibinfo {author} {\bibfnamefont {J.~D.}\ \bibnamefont
  {Pack}},\ }\bibfield  {title} {\bibinfo {title} {Special points for
  brillouin-zone integrations},\ }\href@noop {} {\bibfield  {journal} {\bibinfo
   {journal} {Physical Review B}\ }\textbf {\bibinfo {volume} {13}},\ \bibinfo
  {pages} {5188} (\bibinfo {year} {1976})}\BibitemShut {NoStop}%
\bibitem [{\citenamefont {Baroni}\ \emph {et~al.}(2001)\citenamefont {Baroni},
  \citenamefont {De~Gironcoli}, \citenamefont {Dal~Corso},\ and\ \citenamefont
  {Giannozzi}}]{baroni2001phonons}%
  \BibitemOpen
  \bibfield  {author} {\bibinfo {author} {\bibfnamefont {S.}~\bibnamefont
  {Baroni}}, \bibinfo {author} {\bibfnamefont {S.}~\bibnamefont
  {De~Gironcoli}}, \bibinfo {author} {\bibfnamefont {A.}~\bibnamefont
  {Dal~Corso}},\ and\ \bibinfo {author} {\bibfnamefont {P.}~\bibnamefont
  {Giannozzi}},\ }\bibfield  {title} {\bibinfo {title} {Phonons and related
  crystal properties from density-functional perturbation theory},\ }\href@noop
  {} {\bibfield  {journal} {\bibinfo  {journal} {Reviews of Modern Physics}\
  }\textbf {\bibinfo {volume} {73}},\ \bibinfo {pages} {515} (\bibinfo {year}
  {2001})}\BibitemShut {NoStop}%
\bibitem [{\citenamefont {Garcia}\ \emph {et~al.}(2011)\citenamefont {Garcia},
  \citenamefont {de~Lima}, \citenamefont {Assali},\ and\ \citenamefont
  {Justo}}]{garcia2011group}%
  \BibitemOpen
  \bibfield  {author} {\bibinfo {author} {\bibfnamefont {J.~C.}\ \bibnamefont
  {Garcia}}, \bibinfo {author} {\bibfnamefont {D.~B.}\ \bibnamefont {de~Lima}},
  \bibinfo {author} {\bibfnamefont {L.~V.~C.}\ \bibnamefont {Assali}},\ and\
  \bibinfo {author} {\bibfnamefont {J.~F.}\ \bibnamefont {Justo}},\ }\bibfield
  {title} {\bibinfo {title} {Group {IV} graphene- and graphane-like
  nanosheets},\ }\href@noop {} {\bibfield  {journal} {\bibinfo  {journal} {The
  Journal of Physical Chemistry C}\ }\textbf {\bibinfo {volume} {115}},\
  \bibinfo {pages} {13242} (\bibinfo {year} {2011})}\BibitemShut {NoStop}%
\bibitem [{\citenamefont {Ipaves}\ \emph {et~al.}(2023)\citenamefont {Ipaves},
  \citenamefont {Justo},\ and\ \citenamefont {Assali}}]{ipaves2023aluminum}%
  \BibitemOpen
  \bibfield  {author} {\bibinfo {author} {\bibfnamefont {B.}~\bibnamefont
  {Ipaves}}, \bibinfo {author} {\bibfnamefont {J.~F.}\ \bibnamefont {Justo}},\
  and\ \bibinfo {author} {\bibfnamefont {L.~V.}\ \bibnamefont {Assali}},\
  }\bibfield  {title} {\bibinfo {title} {Aluminum functionalized few-layer
  silicene as anode material for alkali metal ion batteries},\ }\href@noop {}
  {\bibfield  {journal} {\bibinfo  {journal} {Molecular Systems Design \&
  Engineering}\ }\textbf {\bibinfo {volume} {8}},\ \bibinfo {pages} {379}
  (\bibinfo {year} {2023})}\BibitemShut {NoStop}%
\bibitem [{\citenamefont {Cadelano}\ \emph {et~al.}(2010)\citenamefont
  {Cadelano}, \citenamefont {Palla}, \citenamefont {Giordano},\ and\
  \citenamefont {Colombo}}]{cadelano2010elastic}%
  \BibitemOpen
  \bibfield  {author} {\bibinfo {author} {\bibfnamefont {E.}~\bibnamefont
  {Cadelano}}, \bibinfo {author} {\bibfnamefont {P.~L.}\ \bibnamefont {Palla}},
  \bibinfo {author} {\bibfnamefont {S.}~\bibnamefont {Giordano}},\ and\
  \bibinfo {author} {\bibfnamefont {L.}~\bibnamefont {Colombo}},\ }\bibfield
  {title} {\bibinfo {title} {Elastic properties of hydrogenated graphene},\
  }\href@noop {} {\bibfield  {journal} {\bibinfo  {journal} {Physical Review
  B}\ }\textbf {\bibinfo {volume} {82}},\ \bibinfo {pages} {235414} (\bibinfo
  {year} {2010})}\BibitemShut {NoStop}%
\bibitem [{\citenamefont {Ipaves}\ \emph {et~al.}(2022)\citenamefont {Ipaves},
  \citenamefont {Justo},\ and\ \citenamefont
  {Assali}}]{ipaves2022functionalized}%
  \BibitemOpen
  \bibfield  {author} {\bibinfo {author} {\bibfnamefont {B.}~\bibnamefont
  {Ipaves}}, \bibinfo {author} {\bibfnamefont {J.~F.}\ \bibnamefont {Justo}},\
  and\ \bibinfo {author} {\bibfnamefont {L.~V.~C.}\ \bibnamefont {Assali}},\
  }\bibfield  {title} {\bibinfo {title} {Functionalized few-layer silicene
  nanosheets: stability, elastic, structural, and electronic properties},\
  }\href@noop {} {\bibfield  {journal} {\bibinfo  {journal} {Physical Chemistry
  Chemical Physics}\ }\textbf {\bibinfo {volume} {24}},\ \bibinfo {pages}
  {8705} (\bibinfo {year} {2022})}\BibitemShut {NoStop}%
\bibitem [{\citenamefont {Larico}\ \emph {et~al.}(2004)\citenamefont {Larico},
  \citenamefont {Assali}, \citenamefont {Machado},\ and\ \citenamefont
  {Justo}}]{larico2004}%
  \BibitemOpen
  \bibfield  {author} {\bibinfo {author} {\bibfnamefont {R.}~\bibnamefont
  {Larico}}, \bibinfo {author} {\bibfnamefont {L.~V.~C.}\ \bibnamefont
  {Assali}}, \bibinfo {author} {\bibfnamefont {W.~V.~M.}\ \bibnamefont
  {Machado}},\ and\ \bibinfo {author} {\bibfnamefont {J.~F.}\ \bibnamefont
  {Justo}},\ }\bibfield  {title} {\bibinfo {title} {Isolated nickel impurities
  in diamond: A microscopic model for the electrically active centers},\
  }\href@noop {} {\bibfield  {journal} {\bibinfo  {journal} {Applied Physics
  Letters}\ }\textbf {\bibinfo {volume} {84}},\ \bibinfo {pages} {720}
  (\bibinfo {year} {2004})}\BibitemShut {NoStop}%
\bibitem [{\citenamefont {Assali}\ \emph {et~al.}(2006)\citenamefont {Assali},
  \citenamefont {Machado},\ and\ \citenamefont {Justo}}]{assali2006manganese}%
  \BibitemOpen
  \bibfield  {author} {\bibinfo {author} {\bibfnamefont {L.~V.~C.}\
  \bibnamefont {Assali}}, \bibinfo {author} {\bibfnamefont {W.~V.~M.}\
  \bibnamefont {Machado}},\ and\ \bibinfo {author} {\bibfnamefont {J.~F.}\
  \bibnamefont {Justo}},\ }\bibfield  {title} {\bibinfo {title} {Manganese
  impurities in boron nitride},\ }\href@noop {} {\bibfield  {journal} {\bibinfo
   {journal} {Applied Physics Letters}\ }\textbf {\bibinfo {volume} {89}},\
  \bibinfo {pages} {072102} (\bibinfo {year} {2006})}\BibitemShut {NoStop}%
\bibitem [{\citenamefont {Assali}\ \emph {et~al.}(2011)\citenamefont {Assali},
  \citenamefont {Machado},\ and\ \citenamefont {Justo}}]{assali20113}%
  \BibitemOpen
  \bibfield  {author} {\bibinfo {author} {\bibfnamefont {L.~V.~C.}\
  \bibnamefont {Assali}}, \bibinfo {author} {\bibfnamefont {W.~V.~M.}\
  \bibnamefont {Machado}},\ and\ \bibinfo {author} {\bibfnamefont {J.~F.}\
  \bibnamefont {Justo}},\ }\bibfield  {title} {\bibinfo {title} {3d transition
  metal impurities in diamond: electronic properties and chemical trends},\
  }\href@noop {} {\bibfield  {journal} {\bibinfo  {journal} {Physical Review
  B}\ }\textbf {\bibinfo {volume} {84}},\ \bibinfo {pages} {155205} (\bibinfo
  {year} {2011})}\BibitemShut {NoStop}%
\bibitem [{\citenamefont {Haastrup}\ \emph {et~al.}(2018)\citenamefont
  {Haastrup}, \citenamefont {Strange}, \citenamefont {Pandey}, \citenamefont
  {Deilmann}, \citenamefont {Schmidt}, \citenamefont {Hinsche}, \citenamefont
  {Gjerding}, \citenamefont {Torelli}, \citenamefont {Larsen}, \citenamefont
  {Riis-Jensen},\ and\ \citenamefont {{\it et
  al.}}}]{haastrup2018computational}%
  \BibitemOpen
  \bibfield  {author} {\bibinfo {author} {\bibfnamefont {S.}~\bibnamefont
  {Haastrup}}, \bibinfo {author} {\bibfnamefont {M.}~\bibnamefont {Strange}},
  \bibinfo {author} {\bibfnamefont {M.}~\bibnamefont {Pandey}}, \bibinfo
  {author} {\bibfnamefont {T.}~\bibnamefont {Deilmann}}, \bibinfo {author}
  {\bibfnamefont {P.~S.}\ \bibnamefont {Schmidt}}, \bibinfo {author}
  {\bibfnamefont {N.~F.}\ \bibnamefont {Hinsche}}, \bibinfo {author}
  {\bibfnamefont {M.~N.}\ \bibnamefont {Gjerding}}, \bibinfo {author}
  {\bibfnamefont {D.}~\bibnamefont {Torelli}}, \bibinfo {author} {\bibfnamefont
  {P.~M.}\ \bibnamefont {Larsen}}, \bibinfo {author} {\bibfnamefont {A.~C.}\
  \bibnamefont {Riis-Jensen}},\ and\ \bibinfo {author} {\bibnamefont {{\it et
  al.}}},\ }\bibfield  {title} {\bibinfo {title} {The computational 2d
  materials database: high-throughput modeling and discovery of atomically thin
  crystals},\ }\href@noop {} {\bibfield  {journal} {\bibinfo  {journal} {2D
  Materials}\ }\textbf {\bibinfo {volume} {5}},\ \bibinfo {pages} {042002}
  (\bibinfo {year} {2018})}\BibitemShut {NoStop}%
\bibitem [{\citenamefont {Marcondes}\ \emph {et~al.}(2021)\citenamefont
  {Marcondes}, \citenamefont {Santos}, \citenamefont {Miranda}, \citenamefont
  {Rocha-Rodrigues}, \citenamefont {Assali}, \citenamefont {Lopes},
  \citenamefont {Ara\'ujo},\ and\ \citenamefont {Petrilli}}]{marcondes2021}%
  \BibitemOpen
  \bibfield  {author} {\bibinfo {author} {\bibfnamefont {M.~L.}\ \bibnamefont
  {Marcondes}}, \bibinfo {author} {\bibfnamefont {S.~S.~M.}\ \bibnamefont
  {Santos}}, \bibinfo {author} {\bibfnamefont {I.~P.}\ \bibnamefont {Miranda}},
  \bibinfo {author} {\bibfnamefont {P.}~\bibnamefont {Rocha-Rodrigues}},
  \bibinfo {author} {\bibfnamefont {L.~V.~C.}\ \bibnamefont {Assali}}, \bibinfo
  {author} {\bibfnamefont {A.~M.~L.}\ \bibnamefont {Lopes}}, \bibinfo {author}
  {\bibfnamefont {J.~P.}\ \bibnamefont {Ara\'ujo}},\ and\ \bibinfo {author}
  {\bibfnamefont {H.~M.}\ \bibnamefont {Petrilli}},\ }\bibfield  {title}
  {\bibinfo {title} {On the stability of calcium and cadmium based
  ruddlesden–popper and double perovskite structures},\ }\href@noop {}
  {\bibfield  {journal} {\bibinfo  {journal} {The Journal of Materials
  Chemistry C}\ }\textbf {\bibinfo {volume} {9}},\ \bibinfo {pages} {15074}
  (\bibinfo {year} {2021})}\BibitemShut {NoStop}%
\bibitem [{\citenamefont {Kresse}\ and\ \citenamefont
  {Furthm{\"u}ller}(1996)}]{kresse1996efficient}%
  \BibitemOpen
  \bibfield  {author} {\bibinfo {author} {\bibfnamefont {G.}~\bibnamefont
  {Kresse}}\ and\ \bibinfo {author} {\bibfnamefont {J.}~\bibnamefont
  {Furthm{\"u}ller}},\ }\bibfield  {title} {\bibinfo {title} {Efficient
  iterative schemes for ab initio total-energy calculations using a plane-wave
  basis set},\ }\href@noop {} {\bibfield  {journal} {\bibinfo  {journal}
  {Physical Review B}\ }\textbf {\bibinfo {volume} {54}},\ \bibinfo {pages}
  {11169} (\bibinfo {year} {1996})}\BibitemShut {NoStop}%
\bibitem [{\citenamefont {Pakornchote}\ \emph {et~al.}(2019)\citenamefont
  {Pakornchote}, \citenamefont {Ektarawong}, \citenamefont {Alling},
  \citenamefont {Pinsook}, \citenamefont {Tancharakorn}, \citenamefont
  {Busayaporn},\ and\ \citenamefont {Bovornratanaraks}}]{pakornchote2019phase}%
  \BibitemOpen
  \bibfield  {author} {\bibinfo {author} {\bibfnamefont {T.}~\bibnamefont
  {Pakornchote}}, \bibinfo {author} {\bibfnamefont {A.}~\bibnamefont
  {Ektarawong}}, \bibinfo {author} {\bibfnamefont {B.}~\bibnamefont {Alling}},
  \bibinfo {author} {\bibfnamefont {U.}~\bibnamefont {Pinsook}}, \bibinfo
  {author} {\bibfnamefont {S.}~\bibnamefont {Tancharakorn}}, \bibinfo {author}
  {\bibfnamefont {W.}~\bibnamefont {Busayaporn}},\ and\ \bibinfo {author}
  {\bibfnamefont {T.}~\bibnamefont {Bovornratanaraks}},\ }\bibfield  {title}
  {\bibinfo {title} {Phase stabilities and vibrational analysis of hydrogenated
  diamondized bilayer graphenes: A first principles investigation},\
  }\href@noop {} {\bibfield  {journal} {\bibinfo  {journal} {Carbon}\ }\textbf
  {\bibinfo {volume} {146}},\ \bibinfo {pages} {468} (\bibinfo {year}
  {2019})}\BibitemShut {NoStop}%
\bibitem [{\citenamefont {Mouhat}\ and\ \citenamefont
  {Coudert}(2014)}]{mouhat2014necessary}%
  \BibitemOpen
  \bibfield  {author} {\bibinfo {author} {\bibfnamefont {F.}~\bibnamefont
  {Mouhat}}\ and\ \bibinfo {author} {\bibfnamefont {F.-X.}\ \bibnamefont
  {Coudert}},\ }\bibfield  {title} {\bibinfo {title} {Necessary and sufficient
  elastic stability conditions in various crystal systems},\ }\href@noop {}
  {\bibfield  {journal} {\bibinfo  {journal} {Physical Review B}\ }\textbf
  {\bibinfo {volume} {90}},\ \bibinfo {pages} {224104} (\bibinfo {year}
  {2014})}\BibitemShut {NoStop}%
\bibitem [{\citenamefont {Lee}\ \emph {et~al.}(2008)\citenamefont {Lee},
  \citenamefont {Wei}, \citenamefont {Kysar},\ and\ \citenamefont
  {Hone}}]{lee2008measurement}%
  \BibitemOpen
  \bibfield  {author} {\bibinfo {author} {\bibfnamefont {C.}~\bibnamefont
  {Lee}}, \bibinfo {author} {\bibfnamefont {X.}~\bibnamefont {Wei}}, \bibinfo
  {author} {\bibfnamefont {J.~W.}\ \bibnamefont {Kysar}},\ and\ \bibinfo
  {author} {\bibfnamefont {J.}~\bibnamefont {Hone}},\ }\bibfield  {title}
  {\bibinfo {title} {Measurement of the elastic properties and intrinsic
  strength of monolayer graphene},\ }\href@noop {} {\bibfield  {journal}
  {\bibinfo  {journal} {Science}\ }\textbf {\bibinfo {volume} {321}},\ \bibinfo
  {pages} {385} (\bibinfo {year} {2008})}\BibitemShut {NoStop}%
\bibitem [{\citenamefont {Chernozatonskii}\ \emph {et~al.}(2011)\citenamefont
  {Chernozatonskii}, \citenamefont {Sorokin}, \citenamefont {Kuzubov},
  \citenamefont {Sorokin}, \citenamefont {Kvashnin}, \citenamefont {Kvashnin},
  \citenamefont {Avramov},\ and\ \citenamefont
  {Yakobson}}]{chernozatonskii2011influence}%
  \BibitemOpen
  \bibfield  {author} {\bibinfo {author} {\bibfnamefont {L.~A.}\ \bibnamefont
  {Chernozatonskii}}, \bibinfo {author} {\bibfnamefont {P.~B.}\ \bibnamefont
  {Sorokin}}, \bibinfo {author} {\bibfnamefont {A.~A.}\ \bibnamefont
  {Kuzubov}}, \bibinfo {author} {\bibfnamefont {B.~P.}\ \bibnamefont
  {Sorokin}}, \bibinfo {author} {\bibfnamefont {A.~G.}\ \bibnamefont
  {Kvashnin}}, \bibinfo {author} {\bibfnamefont {D.~G.}\ \bibnamefont
  {Kvashnin}}, \bibinfo {author} {\bibfnamefont {P.~V.}\ \bibnamefont
  {Avramov}},\ and\ \bibinfo {author} {\bibfnamefont {B.~I.}\ \bibnamefont
  {Yakobson}},\ }\bibfield  {title} {\bibinfo {title} {Influence of size effect
  on the electronic and elastic properties of diamond films with nanometer
  thickness},\ }\href@noop {} {\bibfield  {journal} {\bibinfo  {journal} {The
  Journal of Physical Chemistry C}\ }\textbf {\bibinfo {volume} {115}},\
  \bibinfo {pages} {132} (\bibinfo {year} {2011})}\BibitemShut {NoStop}%
\bibitem [{\citenamefont {Gercek}(2007)}]{gercek2007poisson}%
  \BibitemOpen
  \bibfield  {author} {\bibinfo {author} {\bibfnamefont {H.}~\bibnamefont
  {Gercek}},\ }\bibfield  {title} {\bibinfo {title} {Poisson's ratio values for
  rocks},\ }\href@noop {} {\bibfield  {journal} {\bibinfo  {journal}
  {International Journal of Rock Mechanics and Mining Sciences}\ }\textbf
  {\bibinfo {volume} {44}},\ \bibinfo {pages} {1} (\bibinfo {year}
  {2007})}\BibitemShut {NoStop}%
\bibitem [{\citenamefont {Hess}(2021)}]{hess2021bonding}%
  \BibitemOpen
  \bibfield  {author} {\bibinfo {author} {\bibfnamefont {P.}~\bibnamefont
  {Hess}},\ }\bibfield  {title} {\bibinfo {title} {Bonding, structure, and
  mechanical stability of 2d materials: the predictive power of the periodic
  table},\ }\href@noop {} {\bibfield  {journal} {\bibinfo  {journal} {Nanoscale
  Horizons}\ }\textbf {\bibinfo {volume} {6}},\ \bibinfo {pages} {856}
  (\bibinfo {year} {2021})}\BibitemShut {NoStop}%
\bibitem [{\citenamefont {Green}(1990)}]{green1990intrinsic}%
  \BibitemOpen
  \bibfield  {author} {\bibinfo {author} {\bibfnamefont {M.~A.}\ \bibnamefont
  {Green}},\ }\bibfield  {title} {\bibinfo {title} {Intrinsic concentration,
  effective densities of states, and effective mass in silicon},\ }\href@noop
  {} {\bibfield  {journal} {\bibinfo  {journal} {Journal of Applied Physics}\
  }\textbf {\bibinfo {volume} {67}},\ \bibinfo {pages} {2944} (\bibinfo {year}
  {1990})}\BibitemShut {NoStop}%
\bibitem [{\citenamefont {Wickramaratne}\ \emph {et~al.}(2015)\citenamefont
  {Wickramaratne}, \citenamefont {Zahid},\ and\ \citenamefont
  {Lake}}]{wickramaratne2015electronic}%
  \BibitemOpen
  \bibfield  {author} {\bibinfo {author} {\bibfnamefont {D.}~\bibnamefont
  {Wickramaratne}}, \bibinfo {author} {\bibfnamefont {F.}~\bibnamefont
  {Zahid}},\ and\ \bibinfo {author} {\bibfnamefont {R.~K.}\ \bibnamefont
  {Lake}},\ }\bibfield  {title} {\bibinfo {title} {Electronic and
  thermoelectric properties of van der waals materials with ring-shaped valence
  bands},\ }\href@noop {} {\bibfield  {journal} {\bibinfo  {journal} {Journal
  of Applied Physics}\ }\textbf {\bibinfo {volume} {118}},\ \bibinfo {pages}
  {075101} (\bibinfo {year} {2015})}\BibitemShut {NoStop}%
\bibitem [{\citenamefont {Novoselov}\ \emph {et~al.}(2016)\citenamefont
  {Novoselov}, \citenamefont {Mishchenko}, \citenamefont {Carvalho},\ and\
  \citenamefont {Castro~Neto}}]{novoselov20162d}%
  \BibitemOpen
  \bibfield  {author} {\bibinfo {author} {\bibfnamefont {K.~S.}\ \bibnamefont
  {Novoselov}}, \bibinfo {author} {\bibfnamefont {A.}~\bibnamefont
  {Mishchenko}}, \bibinfo {author} {\bibfnamefont {A.}~\bibnamefont
  {Carvalho}},\ and\ \bibinfo {author} {\bibfnamefont {A.~H.}\ \bibnamefont
  {Castro~Neto}},\ }\bibfield  {title} {\bibinfo {title} {2d materials and van
  der waals heterostructures},\ }\href@noop {} {\bibfield  {journal} {\bibinfo
  {journal} {Science}\ }\textbf {\bibinfo {volume} {353}},\ \bibinfo {pages}
  {aac9439} (\bibinfo {year} {2016})}\BibitemShut {NoStop}%
\bibitem [{\citenamefont {Garcia}\ \emph {et~al.}(2009)\citenamefont {Garcia},
  \citenamefont {Justo}, \citenamefont {Machado},\ and\ \citenamefont
  {Assali}}]{garcia2009}%
  \BibitemOpen
  \bibfield  {author} {\bibinfo {author} {\bibfnamefont {J.~C.}\ \bibnamefont
  {Garcia}}, \bibinfo {author} {\bibfnamefont {J.~F.}\ \bibnamefont {Justo}},
  \bibinfo {author} {\bibfnamefont {W.~V.~M.}\ \bibnamefont {Machado}},\ and\
  \bibinfo {author} {\bibfnamefont {L.~V.~C.}\ \bibnamefont {Assali}},\
  }\bibfield  {title} {\bibinfo {title} {Functionalized adamantane: building
  blocks for nanostructure self-assembly},\ }\href@noop {} {\bibfield
  {journal} {\bibinfo  {journal} {Physical Review B}\ }\textbf {\bibinfo
  {volume} {80}},\ \bibinfo {pages} {125421} (\bibinfo {year}
  {2009})}\BibitemShut {NoStop}%
\bibitem [{\citenamefont {Chaves}\ \emph {et~al.}(2020)\citenamefont {Chaves},
  \citenamefont {Azadani}, \citenamefont {Alsalman}, \citenamefont {Da~Costa},
  \citenamefont {Frisenda}, \citenamefont {Chaves}, \citenamefont {Song},
  \citenamefont {Kim}, \citenamefont {He}, \citenamefont {Zhou} \emph
  {et~al.}}]{chaves2020bandgap}%
  \BibitemOpen
  \bibfield  {author} {\bibinfo {author} {\bibfnamefont {A.}~\bibnamefont
  {Chaves}}, \bibinfo {author} {\bibfnamefont {J.~G.}\ \bibnamefont {Azadani}},
  \bibinfo {author} {\bibfnamefont {H.}~\bibnamefont {Alsalman}}, \bibinfo
  {author} {\bibfnamefont {D.}~\bibnamefont {Da~Costa}}, \bibinfo {author}
  {\bibfnamefont {R.}~\bibnamefont {Frisenda}}, \bibinfo {author}
  {\bibfnamefont {A.}~\bibnamefont {Chaves}}, \bibinfo {author} {\bibfnamefont
  {S.~H.}\ \bibnamefont {Song}}, \bibinfo {author} {\bibfnamefont {Y.~D.}\
  \bibnamefont {Kim}}, \bibinfo {author} {\bibfnamefont {D.}~\bibnamefont
  {He}}, \bibinfo {author} {\bibfnamefont {J.}~\bibnamefont {Zhou}}, \emph
  {et~al.},\ }\bibfield  {title} {\bibinfo {title} {Bandgap engineering of
  two-dimensional semiconductor materials},\ }\href@noop {} {\bibfield
  {journal} {\bibinfo  {journal} {npj 2D Materials and Applications}\ }\textbf
  {\bibinfo {volume} {4}},\ \bibinfo {pages} {29} (\bibinfo {year}
  {2020})}\BibitemShut {NoStop}%
\end{thebibliography}%

\end{document}